\documentclass[sigconf]{acmart}
\makeatletter
\def\@ACM@checkaffil{% Only warnings
    \if@ACM@instpresent\else
    \ClassWarningNoLine{\@classname}{No institution present for an affiliation}%
    \fi
    \if@ACM@citypresent\else
    \ClassWarningNoLine{\@classname}{No city present for an affiliation}%
    \fi
    \if@ACM@countrypresent\else
        \ClassWarningNoLine{\@classname}{No country present for an affiliation}%
    \fi
}
\makeatother

\usepackage[colorinlistoftodos,prependcaption,textsize=small]{todonotes}
\usepackage{xspace}
\usepackage{listings}
\usepackage{balance}
\usepackage{enumitem}

\newcommand{\system}{\texttt{QO-Advisor}\xspace}
\newcommand{\eat}[1]{}

\newcommand{\stitle}[1]{\vspace{0.5ex}\noindent{\bf #1}}
\newcommand{\mi}[1]{[\textcolor{blue}{[MI:] #1}]}
\newcommand{\zwd}[1]{[\textcolor{blue}{[ZWD:] #1}]}
\newcommand{\paul}[1]{[\textcolor{blue}{[PM:] #1}]}
\newcommand{\aj}[1]{[\textcolor{blue}{[AJ:] #1}]}

\AtBeginDocument{%
  \providecommand\BibTeX{{%
    \normalfont B\kern-0.5em{\scshape i\kern-0.25em b}\kern-0.8em\TeX}}}

\copyrightyear{2022} 
\acmYear{2022} 
\setcopyright{acmlicensed}\acmConference[SIGMOD '22]{Proceedings of the 2022 International Conference on Management of Data}{June 12--17, 2022}{Philadelphia, PA, USA}
\acmBooktitle{Proceedings of the 2022 International Conference on Management of Data (SIGMOD '22), June 12--17, 2022, Philadelphia, PA, USA}
\acmPrice{15.00}
\acmDOI{10.1145/3514221.3526052}
\acmISBN{978-1-4503-9249-5/22/06}

\begin{document}
\fancyhead{}

% \title{\system: Instance-optimized Rule Configurations for SCOPE}
\title{Deploying a Steered Query Optimizer in Production at Microsoft}

% \author{Wangda Zhang$^1$, Matteo Interlandi$^2$, Paul Mineiro$^1$, Shi Qiao$^2$, Nasim Ghazanfari$^2$,\\Karlen Lie$^2$, Marc Friedman$^2$, Rafah Hosn$^1$, Hiren Patel$^2$, Alekh Jindal$^2$}
\author{Wangda Zhang, Matteo Interlandi, Paul Mineiro, Shi Qiao, Nasim Ghazanfari\\Karlen Lie, Marc Friedman, Rafah Hosn, Hiren Patel, Alekh Jindal}

% \affiliation{\vspace{0.5em}$^1$Microsoft Research, $^2$Microsoft\\
\affiliation{\vspace{0.5em}Microsoft\\
  \texttt{qo-advisor@microsoft.com}\hspace{-2em}~
}

\renewcommand{\shortauthors}{Wangda Zhang, Matteo Interlandi, Paul Mineiro, et al.}

\begin{abstract}
Modern analytical workloads are highly heterogeneous and massively complex, making generic query optimizers untenable for many customers and scenarios.
As a result, it is important to specialize these optimizers to instances of the workloads.
In this paper, we continue a recent line of work in steering a query optimizer towards better plans for a given workload, and make major strides in pushing previous research ideas to production deployment. Along the way we solve several operational challenges including, making steering actions more manageable, keeping the costs of steering within budget, and avoiding unexpected performance regressions in production. Our resulting system, \system, essentially externalizes the query planner to a massive offline pipeline for better exploration and specialization. We discuss various aspects of our design and show detailed results over production SCOPE workloads at Microsoft, where the system is currently enabled by default.
\end{abstract}




\maketitle

\section{Introduction}
\label{sec:introduction}

%\rev{Alekh and Matteo}

General purpose query optimization is hitting the end of the road in modern cloud-based data processing systems.
These systems witness a wide variety of highly complex applications that are very hard to optimize globally~\cite{microlearner}, and yet these systems come with generic query optimizers that were built for all users, scenarios, and scales.
Or worse, they are tuned for generic benchmarks like TPC-H or TPC-DS that often do not represent the real customer workloads~\cite{sparkcruise21}.
As a result, these general purpose query optimizers are often far from optimal in their plan  choices for a given customer and a given workload~\cite{howgoodreally}. 
Therefore, there is a need to \textit{specialize} the cloud-based query optimizers to the needs of specific users and applications at hand, and thus optimizing the workloads more effectively, also sometimes referred to as \textit{instance optimization}~\cite{instanceopt-datasys}.

Recent work on instance optimization have proposed to use machine learning to learn from a given user workload different components of a query optimizer that indirectly lead to better query plan choices.
More ambitiously, Neo~\cite{marcus2019neo} proposed to replace the entire query optimizer with a learned one and producing the query plans directly.
Given that replacing an entire query optimizer is not possible in a real system, the follow-up work, BAO~\cite{Marcus2020BaoLT}, proposed to steer the optimizer towards better plan choices by providing rule hints to navigate the search space better for each query.
While BAO considered synthetic workloads on PostgreSQL, our previous work~\cite{steerqo} grounded the ideas in BAO to real-world workloads in industry strength cloud data processing system, SCOPE~\cite{wu2012scope}, illustrating both the challenges and the opportunities in an enterprise workload setting.
In this paper, we continue this line of work and take the above ideas all the way to production deployment at Microsoft. Along the way, we address three sets of challenges to make such a query optimizer operational, as discussed below.

First, prior approaches to steering the query optimizer consider multiple rule hints (as many as 250 rules in the SCOPE query optimizer) at the same time to help the optimizer navigating the search space. The problem, however, is that it is hard to understand or explain what really was wrong and what helped in making the better choice. Thus, neither the system developers can learn about the quality of their query optimizer code bases, nor the service operators explain the impact of the change in case of performance regression, nor the users can build confidence on the robustness of the steering approach.
Second, the experimentation costs involved in prior steering approach were non-trivial. This is because of random sampling in the space of steering hints which when executed in a large scale processing system like SCOPE leads to significant pre-production costs, making it difficult to maintain over time.
And third, steering the query optimizer to newer unseen points in the search space could lead to performance regression, for newer queries and over time. This is because the estimated query costs do not necessarily lead to better plans due to inaccurate cost models~\cite{steerqo}. 
These performance regressions are not acceptable for critical workloads in a running system since users expect a consistent system behavior with predictable performance.

We present \system{} to overcome the above deployment challenges and to steer the optimizer in an explainable, cost-effective, and safe manner. Our key ideas include breaking down the steering process into smaller incremental steps that are easily explainable and reversible.
We introduce a novel model pipeline configuration where a contextual bandit model is used to significantly cut down the pre-production experimentation costs, 
followed by a cloud variability model determining performance improvement or regression.
Essentially, \system{} opens up the core of a query optimizer, i.e., the query planner, which was historically seen as a black box, and externalizes it for better specialization over a customer workload.
Consequently, to the best of our knowledge, \system is the first production deployment of steering a query optimizer.

\vspace{0.5ex}
\stitle{Contributions and Organization.} We make the following major contributions in this paper.

\begin{enumerate}[topsep=0pt]
    \item We present a background to the line of work on steering query optimizer, discussing the current state of the art, the {\color{black}productization} challenges, and our design principles for \system{}. (Section~\ref{sec:steering})
    
    \item We describe the learning principles that we implemented in \system{} to {\color{black}overcome} the {\color{black}productization} challenges. These include using a contextual bandit model to recommend, and a validation model used to accept or reject suggested modification to the query plans. (Section~\ref{sec:training})
    
    \item We discuss several operationalization aspects, including how we integrate with the Azure Personalizer~\cite{agarwal2017making} service, how we deal with plans having non-tree format, and how we featurize such plans. (Section~\ref{sec:operationalization})
    
    \item We show experimental results from production workloads and demonstrate the overall impact that customers can expect, as well as present an in depth analysis of the factors contributing to it. (Section~\ref{sec:experiments})
    
   \item We discuss the lessons we learned over time as we implemented the ideas firstly presented in BAO from a research prototype to a production-ready system. (Section~\ref{sec:lessons})
\end{enumerate}

%\stitle{Organization.}
%The remainder of the paper is organized as follows...
\section{Steering QO: One Year Later}
\label{sec:steering}

%\rev{Matteo, Alekh and Paul}

In this Section, we describe our journey in the steering query optimizer research over the last year.
We first give some background on both the problem that \system tries to address, and the relevant system details from the SCOPE processing engine at Microsoft (Section~\ref{sec:background}).
%we briefly summarize the problem \system tries to address 
Then, we discuss the various challenges we encountered while taking our initial steering optimizer implementation to production deployment (Section~\ref{sec:challenges}). 
Finally, we introduce \system to overcome these challenges (Section~\ref{sec:qoadvisor}), discuss the design principles we followed when implementing the \system (Section~\ref{sec:overview}), and provide an overview of the {\color{black}end-to-end} pipeline involved in the steering process (Section~\ref{sec:pipeline_overview}).

% \vspace{-3ex}
\subsection{Background}
\label{sec:background}

We start with a quick summary of the SCOPE~\cite{scope2021} data processing system at Microsoft, followed by a brief overview of the prior attempt to apply steering optimizer ideas to SCOPE~\cite{steerqo}. 
\system builds upon that work and extends it for production readiness.

\stitle{SCOPE.}
SCOPE is a large scale distributed data processing system~\cite{chaiken2008scope}. It powers
production workloads from a wide range of Microsoft products, processing petabytes of data on hundreds of thousands of machines every day. SCOPE uses a SQL-like scripting language that is compiled into Direct Acyclic Graphs (DAGs) of operators. SCOPE scripts, commonly referred to as \emph{jobs}, are composed as a data flow of one or more SQL statements that are stitched together into a single DAG by the SCOPE compiler.
The SCOPE optimizer is implemented similarly to the traditional
cascades-style~\cite{cascades} query optimizer. As such, it uses a set of rules to transform a logical query plan in a top-down fashion, but it also makes all the decisions required to produce a distributed plan. Examples of such decisions are determining how to partition the
inputs, or selecting the optimal amount of parallelism given the
number of containers available for the job. The number of concurrent containers used by each job is referred to as number of \emph{tokens}
in SCOPE, while the total amount of tokens used for executing a job is called \emph{vertices}.
The SCOPE optimizer estimates the cost of a plan using a combination of data statistics and other heuristics tuned over the years.
Once executed, several metrics are logged by the SCOPE runtime. Common metrics of interests are job \emph{latency}, \emph{vertices count}, and \emph{PNhours} (i.e., the sum of the total CPU and I/O time over all vertices).
Finally, SCOPE provides an A/B testing infrastructure (called \emph{Flighting Service}) which allows to re-run jobs in a pre-production environment using different engine configurations and compare the performance of those configurations with the default one.

\stitle{The importance of recurring jobs.}
More than 60\% of SCOPE
jobs are \emph{recurring}, i.e., periodically arriving template-scripts with different {\color{black}input cardinalities} and {\color{black} filter predicates}~\cite{jyothi2016morpheus, cloudviews}
{\color{black}, but same set of operators. While each execution of a template can have different metrics such as input cardinalities and selectivities,} 
recurring jobs are overall interesting because we can use historical information on previous executions to improve future occurrences.
Because of this, in the remainder of the paper we will focus on recurring jobs.
We will touch upon in Section~\ref{sec:future} how we are planning to remove such limitation in future versions of \system.

\stitle{SCOPE rule-based optimizer.}
There are
256 rules in the SCOPE optimizer. SCOPE rules are divided into 4 categories~\cite{steerqo}: \emph{required} (which must always be enabled to get valid plans), \emph{off-by-default} (which are disabled by default because experimental or very sensitive to estimates), \emph{on-by-default}, and \emph{implementation} rules {\color{black}(mapping logical operators into physical ones)}.
The default optimizer \emph{rule configuration} may return sub-optimal plans for certain job instances or workloads, whereas an instance-optimized rule configuration could return better instance-optimal plans. This, for example, can be achieved by turning on an off-by-default rule, or conversely by turning off an on-by-default or implementation rule (and therefore restricting a part of the optimizer search space).
SCOPE users are aware of this, and in fact over the years have attempted to use \emph{optimizer hints} to override the default query optimizer behavior.
This, however, is a manual task which requires both deep knowledge of SCOPE internals, as well as tedious experimentation through trials and errors.
Despite this, on a daily basis we see that up to 9\% of jobs have hints overriding the default configuration of the optimizer.
Applying state-of-the-art techniques such as BAO~\cite{Marcus2020BaoLT} is non-trivial, and it will not work because the configuration space is just too big: $2^{256}$ in SCOPE vs just $48$ considered in BAO.

\stitle{Towards instance-optimized rule configurations.}
To overcome the above problem, in \cite{steerqo} we have explored a first step towards automating this manual experimentation task. 
We analyzed large production workloads from SCOPE and introduced two new concepts to apply the steering ideas in such a system:
%Both \cite{steerqo} and this work are based on two important concepts:
\begin{description}
    \item[Rule Signature.] For each plan, we extended the SCOPE optimizer to return a bit vector specifying which rules directly contributed to it. For example, if only the first and the second rule were used during the optimization, then the rule signature will be 1100000000.
    \item[Job Span.] Given a job, we compute a set containing all rules which, if enabled or disabled, can affect the final query plan. Intuitively, the span of a job contains all the rules that can lead to some change in the final optimized  plan. The job span allows to narrow down the exploration of rule configurations by skipping the unworthy ones. 
    The job span is computed heuristically, as described in \cite{steerqo}.
\end{description}

Given the rule signature for a job, and its span, in \cite{steerqo} we showed that {\color{black}it} is possible to generate instance-optimized rule configurations returning better estimated cost and runtime metrics (up to $90\%$ latency improvement) than using the default rule configuration.
The heuristic used in \cite{steerqo} for the configuration search is as follows:
\begin{enumerate}
    \item For each job, randomly sample $1000$ configurations over that job's span following a uniform distribution;
    \item Recompile all generated samples and confront the new estimated cost returned by the optimizer against the ones from the default rule configuration;
    \item For all configurations with better cost estimates, pick the $10$ more promising one, and flight them against the default configuration for validating that better cost estimates translate into better runtime metrics;
    \item Among all the flighted configurations, pick the one with the best runtime metrics improving over the default configuration (if it exists), and apply such configuration to the next occurrence of the recurring job.
\end{enumerate}

In this paper, we will close the loop, and describe how we are able to achieve fully automated instance-optimized rule configurations of SCOPE jobs with \system.

\subsection{Productization Challenges}
\label{sec:challenges}

Over the past one year, we have been trying to deploy in production the heuristics of~\cite{steerqo} for steering the SCOPE query optimizer. However, while doing this, we faced several challenges.
\begin{description}
    \item[Difficult to debug.] Instance-optimized rule configurations are generated using uniform sampling. This provides no real insight on why a configuration is better than another. Consequently, when an IcM (Incident Management ticket) is raised (e.g., because of some unexpected performance regressions) debugging is almost impossible.
    \item[Expensive to maintain.] The amount of resources required to maintain this feature in production is not trivial. In fact, for each recurring job we recompile $1000$ different rule configurations, as well as flight (re-run in pre-production environment) $10$ of them, which can quickly become very expensive.
    \item[Performance regressions are hard to catch upfront.] There is no principled way of detecting regression or bad configurations upfront. The only guard we have are the estimated costs from the SCOPE optimizer after recompilation (whose reliability is  well known to be lacking~\cite{siddiqui2020cost}) and flighting (which does not consider the variance inherent in the clusters, unless we repeat each flighting run several times). 
\end{description}

% Next we are going to give a high level overview of \system's approach, and how it is able to address all of the above challenges.
% \system is currently running in production and enabled by default {\color{black}on some the SCOPE workloads}.

% \vspace{-2ex}
\subsection{Introduction of \system}
\label{sec:qoadvisor}

We now give an overview of the \system, which addresses the above challenges in SCOPE-like big data query optimizers. Similar to index recommendations by an index advisor~\cite{ai-meets-ai,cophy,sql-server-advisor}, the goal of \system is to recommend better search paths, via rule hints, for a query plan. The difference, however, is that while index recommendations are applied offline and appropriate indexes are created a priori, the search space recommendations are applied online during the query optimization process itself. Therefore, to keep this process lightweight, \system does all the heavy lifting of generating the appropriate rule hints for different recurring job templates in an offline pipeline.
The idea is to leverage the massive past telemetry to explore better query plans in an offline loop, and integrate the recommendations to steer the optimizer to those plans in the future.
Figure~\ref{fig:overview} shows how the \system pipeline sits next to the SCOPE engine and helps {\color{black}steer} its search space.

\system opens up a core query optimizer component, namely the query planner, to be explored and improved upon externally via the use of past telemetry and offline compute resources.
This is a major shift from traditional query planning 
and allows {\color{black}one} to think differently. For example, we can 
consider more expensive and more exhaustive search algorithms which one cannot afford during query optimization. We can
observe the actual execution costs seen in the past as compared to relying purely on the estimated costs. We can
do more validation offline to anticipate any unexpected query plan behavior and even improve upon it before actually deploying it to future queries.
To top it all, \system integrates back seamlessly with the query optimization of future queries in an automated loop, without having users to worry about or even notice the externalized query planning component.
Thus, \system introduces a new way to think about query planning that helps scale query optimizers to the complexity and challenges that we see in modern cloud workloads~\cite{microlearner}.

\subsection{Design Principles}
\label{sec:overview}

As mentioned above, the \system is a pipeline of tasks that is recurrently triggered every day. 
\system pipeline takes as input historical metadata for a given day, and return a list of job template identifiers and rule hint pairs.
These pairs are consumed by the SCOPE optimizer such that, every time a job matching one of the template identifier is found, the provided rule hint is used at compile time to steer the query optimizer.

\stitle{Single rule flip.}
The first major difference compared to previous approaches is that \system does not provide full rule configurations, but it only amends the default SCOPE rule configuration by turning on or off a single rule at a time.
We made this design decision for several reasons: (1)~the search space is much smaller with single rule flips and therefore easier to manage and less expensive to maintain; (2)~in case of performance regression, one single rule change is far more easier to manage/revert compared to arbitrary changes to the rule configurations; (3)~this simpler approach is easier to control in production settings, therefore better for building confidence with the product teams.

\stitle{From naive to informed experimental design.}
Uniform sampling over the configuration space generated by the job span is expensive to maintain because it requires the generation of several configurations, the majority of which are discarded.
In \system we provide a more principled approach whereby we train a contextual bandit model over features extracted from the historical runs, as detailed in Sections~\ref{sec:training} and ~\ref{sec:operationalization}.
This allows us to decrease the number of required re-compilations, as well as flighting runs. 
%\mi{Wanga, Paul: Can we also say something like in case of regression, we can look at the features and return an explanation of why that configuration was returned?}

\stitle{Learning over estimates rather than runtime metrics.}
Learning over query execution times, as in BAO, is hard at SCOPE scale. SCOPE daily executes 100,000s of jobs spanning 100,000s of machines, whereby: {\color{black} (1) re-executing even a small fraction of them can be prohibitively expensive; and (2) testing configurations without any (although imprecise) guardrail introduces the chance of executing plans with orders of magnitude worse runtime metrics, which, at SCOPE scales, means a large waste of resources.} {\color{black}Therefore} we decide to learn rule configurations over the estimated costs output by the SCOPE optimizer.
While this approach is not ideal in terms of accuracy (since we are learning to improve estimated costs rather than {\color{black}real ones}), still it allows us to run at SCOPE scale.

%\stitle{Flighting used only to avoid regression.}
\stitle{Models pipeline for avoiding regression.}
Our learned model suggests new rule configurations by predicting the eventual improvement on estimated costs. The question on how to avoid regressions is still open, since (predicted or actual) estimated costs can be off compared to the real execution runtime, as previously suggested in Section~\ref{sec:challenges}.
To address this, 
in \system we decide to resort to a model pipeline whereby a validation model is applied after the contextual bandit model.
The goal of the second model is to make sure that the selected rule configuration will not create performance regressions after execution. 
To gather runtime measurements, we use flighting. However we flight only a small number of jobs, and under a tight budget.
The flighting budget is proportional to how strictly we want to avoid performance regressions.

\subsection{Pipeline Overview}
\label{sec:pipeline_overview}

\begin{figure}[t]
\includegraphics[width=0.48\textwidth]{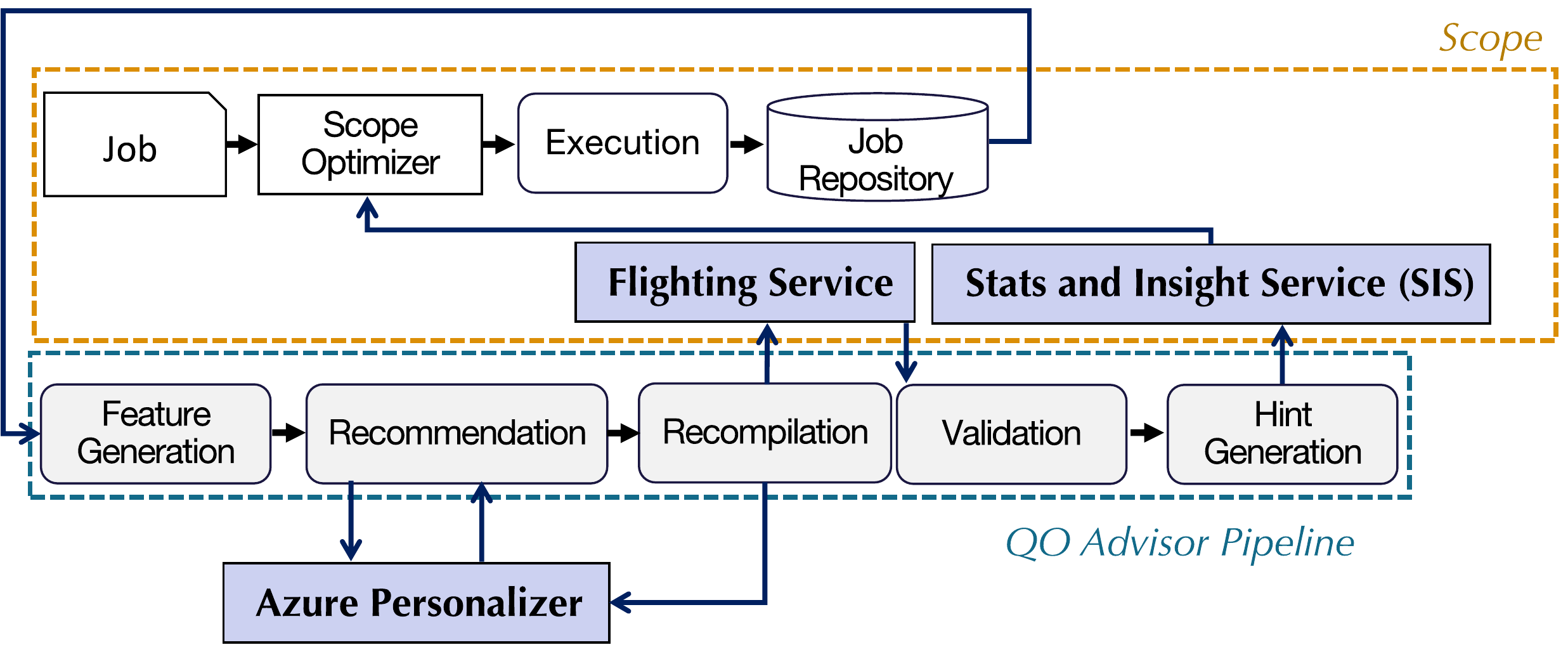}
\vspace{-5ex}
\caption{Overview of the \system pipeline and its integration with SCOPE. The pipeline is composed of five tasks (Feature Generation, {\color{black}Recommendation, Recompilation, Validation} and Hint Generation) and it uses three services: Azure Personalizer, SCOPE Flighting, and Stats and Insight Service.}
\label{fig:overview}
\vspace{-3ex}
\end{figure}

Figure~\ref{fig:overview} provides a sketch of the \system pipeline and its integration with SCOPE.
The \system pipeline is executed offline  and daily. It takes as input historical information on job execution for a specified date, and produces a list of pairs (job template, hints) which are loaded into SCOPE's optimizer through the Stats and Insight Service (SIS)~\cite{jindal2019peregrine}.
From the historical runs metadata, the pipeline first generates a set of features, and then feeds them into a Recommendation task which uses Contextual Bandit~\cite{langford2007epoch} to suggest up to one rule hint for each job.
These rule hints are then passed to the SCOPE optimizer for a Recompilation run.
Jobs are recompiled such that: (1)~we can catch compilation errors due to the new rule settings upfront; and (2)~we can get a new estimated cost for the plan.
Jobs with worse estimates are pruned, while for the remaining jobs we run an A/B test through the Flighting Service.
The output of flighting is then fed into a Validation task that applies a linear regression model.
This provides a guard over performance regressions, given that it is crucial for production workloads~\cite{perfguard}.
The job templates (and related hints) that pass validation, are then written into a file by the Hint Generation task, and loaded into SIS, which, in turn, loads them into the SCOPE optimizer such that the generated hint is applied to the next occurrence of the job template. 

Next we describe \system's training tasks in the following section. Thereafter, in Section~\ref{sec:operationalization}, we will provide insights on \system's operationlization and integration with SCOPE. 
\section{Learning Principles}
\label{sec:training}

%\rev{Paul, Wangda, Rafah}

In this section we describe how \system learns to steer the SCOPE optimizer. Below we first give a short background to contextual bandits and then describe our formulation for steering the optimizer search space.

%\mi{One thing that I think is missing here or in section 4 is how we generated the training dataset. }
%\zwd{the CB model is actually learning online, Sec 4.2 explains its training set. The validation model is supervised learning using train/test data split in the end of Sec 4.3. }
%\mi{This section should contain the main meat of the paper. I think that all the implementation details should go into Section 4, so here we can stay a bit abstract (e.g., we can mention Personalizer integration in Section 4 and just explain how the RL model works).}

\subsection{Introduction to Contextual Bandits}
\label{subsec:cbintro}

%\zwd{Instead of saying CB is an extension of supervised learning, can we say CB is a subset of RL?} \paul{I'd rather not.  It's only a subset of RL in a manner that a theoretician would recognize.  For this audience it is better conceptualized as an extension to supervised learning.}

Contextual bandits {\color{black}(CBs)} are an extension of supervised learning where the feedback is limited to the actions made by the learning algorithm~\cite{langford2007epoch}.  For optimization problems where enumeration is impractical, {\color{black}CB} is a useful abstraction because they limit the evaluation burden to the actions made by the learning algorithm.  Specifically, in our problem setting, the burden of computing the quality of all plausible {\color{black}rule configurations} implies supervised learning has a very high overhead, but {\color{black}CB} learning has low overhead as only the {\color{black}rule configuration selected} by the learning algorithm will need to be evaluated.

{\color{black}CBs} achieve statistical guarantees similar to those of supervised learning by constructing an experimental design and then reducing it to supervised learning~\cite{agarwal2014taming,foster2021adapting}. However, the use of supervised learning as a subroutine implies {\color{black}CB} learning inherits {\color{black}some of the }issues of supervised learning. In particular, query plan representations can have a large impact on performance.  We discuss issues of representation later in this section.

The experimental designs induced by {\color{black}CB} algorithms are randomized; informally, good actions become increasingly more likely under the experiment design as more data accumulates, but other actions still have some likelihood despite their poor historical performance. Specifically to our setting, a {\color{black}CB} algorithm will initially choose {\color{black}rule configurations uniformly-at-random}, but as historical data accumulates it will play (what the supervised learning model indicates is) the best {\color{black}rules configuration} with increasing frequency.  Although the guarantees of {\color{black}CB} learning have the same  dependence on data volume as supervised learning, the effective number of data points is {\color{black}proportional, in the worst case, to} the number of {\color{black} possible} actions per example.  This motivates our strategy for limiting the number of actions per example discussed later in this section.

These aspects of {\color{black}CB} learning (limited evaluation requirements and informed randomization) are the basis for the improvements over our previous approach~\cite{steerqo}.

\subsection{Contextual Bandit Formulation}
\label{sec:context}

In {\color{black}CB}, the learning algorithm repeatedly receives a \emph{context} and \emph{action-set} pair $\left(x, \{ a_i \}\right)$, it chooses an action $a_i$, and then receives \emph{reward} $r$.  To formulate query optimization as a {\color{black}CB problem}, we must specify the context, the actions, and the rewards.

\vspace{0.2cm}
\stitle{Recommendation.} The Recommendation task consumes information available after the default compilation and optionally recommends an alternative compilation, as shown in Figure~\ref{fig:overview}.  For this component we use the fractional reduction in (optimizer) estimated cost as the reward.  Rewards with extreme dynamic ranges can cause problems for existing learning algorithms, so we use clipping to mitigate the effects of outliers when training. %(but not when reporting results).

The specification of the action set $\{a_i\}$ leverages the concepts of rule signature and job span introduced in Section~\ref{sec:background}.  The main statistical issue is to limit the number of actions per example in order to control the amount of data required for learning to converge.  With $B$ possible bits in the rule signature, there are na\"ively $2^B$ possible actions corresponding to each rule signature.  This is intractably large so we instead only consider the rule signatures which differ in the job span.  If there are $S$ bits in the job span for a query, then there are $2^S$ possible actions corresponding to possible settings of the span bits.  For this work, we limited the action space to single bit deviations from the default query plan (both to reduce data requirements for learning and for the reasons in Section~\ref{sec:overview}).  Thus the number of possible actions is $(1 + S)$, corresponding to either changing nothing (1) or flipping a single bit in the span ($S$). Empirically, $S$ is on average 10 but with a long tail distribution~\cite{steerqo}, so having actions scale linearly with $S$ ensures tractability.
% See figure 2c in \cite{steerqo}

For the context, any information which is known after the initial default compilation can in principle be utilized. We initially attempted to compute features directly from the logical query plan via properties of the DAG, but this was not effective.  Instead we found representing the logical query by the job span itself to be more effective.  In other words, the complete set of bit positions in the job span provides valuable and concise information about which bits can be flipped to improve the result, especially when interacted to create second and third order co-occurrence indicators.  Beyond these features, we found representing some properties of the input data streams (e.g., row count) provided marginal improvement.  Further details are provided in Section~\ref{sec:operationalization:recommend}.

\vspace{0.2cm}
\stitle{Validation.} The Validation component consumes information available after a single flighting run of the query plan and decides whether to accept or reject the modified plan.  Because (1) there are only two actions, and (2) the reward of reject is known (relative change is 0), we treat this a supervised learning problem and utilize a model trained on a fixed dataset.  Further details are in Section~\ref{sec:operationalization:validate}.

% Example of context.
% \begin{lstlisting}
% {
%  "Span": {"35": 1, "40": 1, "70": 1, "71": 1, "72": 1, "73": 1, "77": 1, "85": 1, "86": 1},
%  "EstTotalCost": {"EstTotalCost_2": 0.963289875626747, "EstTotalCost_3": 0.03671012437325305},
%  "TotalContainers": {"TotalContainers_3": 0.0, "TotalContainers_4": 1.0},
%  "Cardinality": {"Cardinality_3": 0.2110675119042167, "Cardinality_4": 0.7889324880957833},
%  "RowLength": {"RowLength_9": 0.8740619123277852, "RowLength_10": 0.12593808767221482},
%  "_multi": [
%      {"Action": {"RuleID": "40", "RuleCategory": "on-default"}},
%      {"Action": {"RuleID": "86", "RuleCategory": "on-default"}},
%  ],
% }
% \end{lstlisting}

\section{Operationalization}
\label{sec:operationalization}

%\rev{Matteo and Wangda}

%\mi{This section should summarize all the things related to deployment and implementation. We can also add here all the tricks that we had to do, e.g., Scope jobs contains multiple queries (meaning each job is a DAG with multiple roots) and we had to introduce a dummy root aggregating all features, or we flight based on cost estimates (ascending) and tiering.}

In this section we will provide insights on the \system implementation.
We have implemented the \system pipeline entirely in Python. Additionally, the \system interface with external components, namely SCOPE, Flighting Service, SIS and Azure Personalizer~\cite{agarwal2017making}, use a mix of Python and Powershell commands.

We run the \system pipeline daily from a set of Windows machines. \system is currently deployed
for {\color{black}some of SCOPE workloads} running on multiple SCOPE clusters, where it is enabled by default, i.e., every job submitted  {\color{black}within} those workloads gets search space recommendations produced by the \system pipeline. We are currently investigating further integration with automation services such as Azure Data Factory (ADF)~\cite{adf}. 

\system takes as input a denormalized \emph{view} of the workload aggregating all compile time as well as execution information of jobs run on a given date~\cite{jindal2019peregrine}. This view file is generated automatically by an ADF pipeline and is shared across different learned components in SCOPE~\cite{cloudviews,jindal2019peregrine,autotoken,siddiqui2020cost}. 
It contains information such job identifier, name, a description of the job plan, as well as optimizer (e.g., estimated cardinalities, estimated cost) and runtime statistics (e.g., latency, PNhours, actual row counts).
We defer to Section~\ref{sec:operationalization:feature} below for a detailed description on the features available from the above view file, as well as how they are used by our machine learning models in \system.
This denormalized view for a given date is typically available on COSMOS~\cite{scope2021} (the big data platform at Microsoft) within 3 days. This means that \system is triggered over jobs executing not earlier than 3 days from the current date.

\system pipeline is composed of five tasks, as depicted in Figure~\ref{fig:overview}, namely Feature Generation, rule Recommendation with Recompilation, plan Validation, and the final Hint Generation. 
Below, we provide implementation details for each of these tasks.

\subsection{Feature Generation}
\label{sec:operationalization:feature}

In the first step of the \system pipeline, starting from the denormalized workload view, we run a sequence of SCOPE jobs to (1)~generate the span for all jobs, (2)~return all features necessary for the training of the downstream models.

\stitle{Job span generation.} We use the same algorithm illustrated in~\cite{steerqo} without any further modification. Briefly, the span algorithm implements an heuristics search whereby {\color{black} only new rules having effect on the final plan can be discovered. Specifically, for each job we start from the original rule configuration, and we turn on all the off-by-default rules, while we turn off all the on-by-default and implementation rules that appear in the original rule signature.}
We then pass this new rule configuration to the SCOPE optimizer for a recompilation pass. {\color{black} Since some rules that were previously used (and on) are now off, and some other rules that instead were previously off are now available to the optimizer, after recompilation we may have a rule signature with new used rules.}
We then again turn off all the newly used (on-by-default, off-by-default or implementation rules) and run the new configuration through the optimizer. This process is repeated until we reach a fix-point {\color{black}(i.e., no new rule is added to the signature, or the recompilation fails).}
% for each job, rules are iteratively turned off such that new rules having effect on a final plan can be discovered. Starting from the original rule configuration for a job, we turn on all the off-by-default rules, while we turn off all the on-by-default and implementation rules that appear in the rule signature. We then pass this new rule configuration to the SCOPE optimizer for a recompilation pass. After recompilation, we have a new rule signature with eventually new rules used. We then again turn off all the newly used (on-by-default, off-by-default or implementation rules) and run the new configuration through the optimizer. This process is repeated until we reach a fix-point. 
All jobs that have an empty span (i.e., the heuristics cannot find any modification to the default configuration) are not further considered. % by \system.

\stitle{Feature aggregation.}
An interesting aspect of SCOPE is that each job executes a \emph{script} which can contain one or more queries.
Therefore, for each job, the optimized plan is a DAG of operators (instead of a single tree, as is common for relational database plans), with one or more output nodes, one for each resulting dataset.
Each output node can be seen as the root of a tree, whereby the SCOPE optimizer and runtime generates some statistics (features) per tree, while other statistics are generated per script. 
Examples of these per-tree features are row counts, estimated cardinalities and average row length.
Examples of job-level statistics are PNhours, latency and total number of used containers.
Another interesting aspect of having DAGs instead of trees is that each job could be part of different templates, since template information are generated per-query tree and not per-job script.
Finally, since \system executes at the job granularity (and similarly hints can be provided to the SCOPE optimizer at the job level) we found that there is a disconnect between how the features are generated by SCOPE and the format that is needed for \system to operate properly.

To solve the above challenge, from the raw features and plans contained in the view file, we transform DAGs into trees by generating a \emph{super root node}. Super root nodes aggregate all the features and information contained in the sub-query trees composing a job.
Features are aggregated in two different ways: (1)~following their semantics, e.g., we do {\sc sum} over estimated cardinalities, and {\sc avg} over average row length; (2)~for plan-level features as well as for template-level  features, we just get the {\sc min}.
Table~\ref{tab:features} summarizes all the features generated at this step of the pipeline, and the related aggregate function.
After this step, all features are job level and properly aggregated. We then attach the span information and feed the result to the Recommend task, which we describe next. %Next, we describe the recommend task below.

\begin{table}[t]
\centering
\caption{Features generated for each job, how they are aggregated, and their source. For each feature we also highlight whether it is a job-level (J) or a query-level (Q) feature. {\color{black} Note that for job-level features we always use {\sc min} as aggregate function, since all queries within the same job have the same value for those features.}}
\label{tab:features}
\small{
\begin{tabular}{|l|c|c|c|} 
 \hline
 Feature & Aggregation & Source & Level\\
%  & Function & &\\
 \hline\hline
 Normalized Job Name & {\sc min} & Job Metadata & J\\ 
 \hline
 Rule Signature & {\sc min} & Optimizer & J\\
 \hline
 Latency & {\sc min} & Runtime Statistics  & J \\
 \hline
 Estimated Cost & {\sc min} & Optimizer & J \\
 \hline
 Query Template & {\sc min} & Job Metadata & Q \\
 \hline
 Total Number of Vertices & {\sc min} & Runtime Statistics & J \\
 \hline
 Estimated Cardinalities & {\sc sum} & Optimizer & Q\\
 \hline
 Bytes Read & {\sc sum} & Runtime Statistics & Q\\
 \hline
 Maximum Memory Used & {\sc min} & Runtime Statistics & J\\
 \hline
 Average Memory Used & {\sc min} & Runtime Statistics & J\\
 \hline
 Average Row Length & {\sc avg} & Optimizer & Q \\
 \hline
 Row Count & {\sc sum} & Optimizer & Q\\
 \hline
 PNHours & {\sc min} & Runtime Statistics & J \\
 \hline
\end{tabular}
}\vspace{-3ex}
\end{table}

\subsection{Rule Recommendation}
\label{sec:operationalization:recommend}

As formulated in Section~\ref{sec:context}, we use a {\color{black}CB} model to recommend \emph{rule flips} (i.e., turn off an on-rule or turn on an off-rule) for a specific job. More specifically, each action flips a rule in the job span. We featurize the actions using the rules id and rules category information as introduced in Section~\ref{sec:background}.
The context includes the features of Table~\ref{tab:features} generated from the previous step of the \system pipeline, plus the job span. 
% Briefly, these features come from two sources: (1) the aggregated job metadata features such as estimated cardinality, average row length, and total number of containers, which reflects the data set information of a specific job, and (2) information generated by the optimizer including the complete job span (tunable rules and requried rules) of a specific job, and the estimated cost of the optimizer in default rule configurations.
The reward is the relative change in the optimizer's estimated cost, i.e. the ratio between estimated cost in the default setting over the estimated cost after re-compilation with a recommended rule flip. The contextual bandit optimization is to maximize the reward: if a rule flip leads {\color{black} to a} higher reward, the estimated cost after recompilation becomes smaller, so that a plan with lower estimated cost is found. We clip the range of the ratio so that extreme values do not overly skew the recommendation model.
{\color{black} Currently we heuristically clip any range greater than 2.0, i.e., we clip any plan that is more than 2$\times$ the baseline.}
The clipped delta is then fed back to the contextual bandit model as the reward.

For {\color{black}CB} learning we use an off-policy learning approach~\cite{wang2017optimal}, where we gather reward information using the uniform-at-random policy, but for the subsequent steps we act using the learned contextual bandit policy.  This accelerates learning by inducing a maximally informative training dataset, at the cost of doubling the number of rule configurations test compilations. Since job recompilations are relatively inexpensive this is an acceptable trade-off.

In our implementation, we use the Azure Personalizer service~\cite{agarwal2017making} to generate rule recommendations. Azure Personalizer uses the contextual bandit approach introduced in Section~\ref{subsec:cbintro} to predict and learn under a specific problem scenario. Azure Personalizer allows for better development in our scenario compared to ad-hoc solution, thanks to three main advantages: (1) Azure Personalizer automatically handles all aspects of model management, fault-tolerance and high availability; (2) it logs with high fidelity so that we can counter-factually evaluate policies; and (3) it adheres to all Azure compliance and security settings.

% \subsection{Recompilation}

% The output of the previous step has two rule flips for each job: a random rule flip used for training the Personalizer, and a recommended rule flip for flighting and validation.
% In this step we recompile all such jobs with the random rule flip to generate the rewards so that the personalizer can learn from it.

% theoretically we don't need the second recompile with recommended rule flips. This current implementation is mainly due to a limitation in personalizer api.
% We can also recompile jobs using the recommended rule flip to get an updated estimated cost as well as validating that the suggest rule flip generate an admissible rule configuration, e.g., the SCOPE optimizer can effectively compile the jobs without errors. For all successful re-compilations, the jobs become the candidates for the Flighting Service.

\subsection{Validation}
\label{sec:operationalization:validate}

% \mi{Need expand once we are done with~\ref{sec:training}}
% \mi{Here we put everything related to validating flighting results and catching regressions. We can start with showing the variance plots and explain why we need a validation step.}

In the previous rule Recommendation task, we evaluated the rule flips (actions) generated by the {\color{black}CB} model. However, we found that using estimated costs alone could sometimes lead to performance regressions. This is mainly due to the variability inherent in the cloud computing system which makes cost estimation incredibly hard~\cite{siddiqui2020cost}. Interestingly though, we found that the variability problem is more severe for the job latency metric and lesser so for the PNhours metric.
While we will further discuss this finding in Section~\ref{sec:experiments:variance}, where we run A/A tests to evaluate the variance in SCOPE clusters, in summary we found that since PNhours is computed as the sum between CPU time and I/O time, the variablity of I/O time across A/A runs is bounded as data read and data written remain constant.
From the above observations we made two conclusions:
\begin{enumerate}
    \item The variance in the cluster is so high that it makes hard to infer improvements in the runtime reliably and programmatically, without having to execute each job several times.
    \item The total amount of data read and data written are good predictors for whether a rule flip introduces improvement or not. Intuitively, if with the new configuration a job read and write less data, this will likely translate into better runtime.
\end{enumerate}

In order to gather information about data read, data written and other indicators, we use the SCOPE Flighting Service to test the rule flips in a pre-production environment. Next we describe how we use the flight results to validate the recommended actions that are output from the contextual bandit model.

\stitle{Flighting.}
Given a list of jobs and their related recommended actions, we use the estimated costs (over the recompiled plans with the rule configurations embedding the flips suggested by the model) to heuristically select which jobs to flight. Fighting jobs is in fact not just expensive, but it is the larger source of resource and time consumption for \system.
Because of this, we have a limited budget of machines that can be concurrently used for flighting, and we set thresholds on (1) the maximum flighting time for each job (24 hours); (2) the total time budget for flighting; and (3) the delta between the estimated cost from recompilation, and the default one.
Additionally, we do not flight all jobs that reach the {\color{black}Recompilation} step, but instead we flight one representative job per template (picked randomly). 
The intuition here is that jobs with the same template are composed by the same queries, and therefore it is not necessary to flight all of them because they have the same plans and eventually the same rule flip.
Finally, we flight jobs with lower estimated costs first, such that if we finish the total time budget, we are still able to learn and provide some suggestion even if we are not able to complete the flighting process in full.

Given our limited budget on the number of jobs that can be concurrently flighted, we interact with the SCOPE Flighting Service through a job queue of fixed size. 
For each job being flighted, the service can {\color{black}return} several different outcomes: (1) failure (e.g., the job information or the input data expired); (2) timeout; (3) filtered (e.g., if the job belongs to certain classes of jobs that are not supported by the Flighting Service); (4) success. 
Only the jobs/templates that are flighted with success are passed to the next step in the pipeline.
{\color{black}While flighting introduces some overhead cost for \system, this cost is one-time for each job, and amortized over time.
The cost is also bound by the available budget, which can be tuned based on how aggressive we want to be in production.}
In Section~\ref{sec:future} we will touch on how we are planning to improve the flighting process for increasing the throughput.

\stitle{Validation model.}
% \zwd{we can also call this a binary classifier for deciding whether the job is put into STP, using flighting results as features.}
The flighting results go over a Validation step for catching possible regressions before running the new rule configuration in production, as suggested in Section~\ref{sec:training}.
In this step we run a linear regression model that learns the PNhours delta given the per job total \emph{DataRead} and \emph{DataWritten features} returned by flighting.
Again, the intuition here is that if a job reads and writes less data, then it is likely that the PNhours will be reduced (despite the variance inherent in the cluster).
In order to train this supervised validation model, we flight a random subset of the jobs over a period of 14 days to gather a data set of flighting results. The data points are indexed by their timestamps, so that we can split the dataset by date to generate a training set (e.g., data in week0) and a testing set (data in week1), in order to test whether the trained model can generalize to other dates temporally.

The output of the {\color{black}Validation} model is then compared against a pre-determined safety threshold such that only when the PNhours delta is below this threshold we are actually confident with running the chosen action in production, without causing significant regressions.
The threshold can be increased or decreased based on how much aggressive we want be. At the moment, for the workloads we are currently running in production, this threshold is set to -0.1, meaning that, for a particular job, if the expected reduction in PNhours is at least 10\%, the job passes the validation.

% \vspace{-1ex}
\subsection{Hint Generation}
\label{sec:operationalization:hint}
In this final step, we gather the validated (job templates, new rule configuration) pairs and then explode them by applying the same configuration to all jobs belonging to the same template.
The output is saved to a file in the Stats and Insight Service (SIS) pre-defined format, and this file is uploaded to the SIS.
SIS makes deploying models and configurations in SCOPE easier as it manages versioning and validates the format before installing them in the SCOPE optimizer~\cite{jindal2019peregrine}.
In our case, the SIS picks the file containing jobs and related rule configurations, and applies them as hints to the SCOPE optimizer every time a new instance of the same job template is submitted in the future.

\section{Experiments}
\label{sec:experiments}

%\rev{Wangda and Matteo}

%\paul{I see RL abbrevation used alot.  I would prefer we just talk about contextual bandit, using CB as an abbrevation.  Perhaps less sexy, but more accurate, while still intimidating to those unfamiliar with machine learning.}

In this Section, we present several 
results and insights from one of the production SCOPE workloads, in a pre-production environment.
Based on these results, the \system was enabled by default for this workload beginning November 2021.
Throughout this section, we focus on answering the following questions that made the above deployment possible:
\begin{enumerate}
    \item Which runtime metrics are best suited for measuring success? (Section~\ref{sec:experiments:variance})
    \item Can we reliably use compile-time-only information to predict runtime outcomes?  (Section~\ref{sec:experiments:estimates})
    \item Can we reliably predict future runtime outcomes from a single flight run? (Section~\ref{sec:experiments:validate})
    \item Do we improve over default query plans in aggregate? (Section~\ref{sec:experiments:flight})
    \item What is the impact on query performance in the worst case? (Section~\ref{sec:experiments:distribution})
    \item Is our biased search using contextual bandit more effective than a uniformly-at-random baseline? (Section~\ref{sec:experiments:recompile})
\end{enumerate}

\subsection{Metrics}
\label{sec:experiments:variance}
% Why do we care about PN hours

\begin{figure}[t]
\vspace{-4ex}
\includegraphics[width=\linewidth]{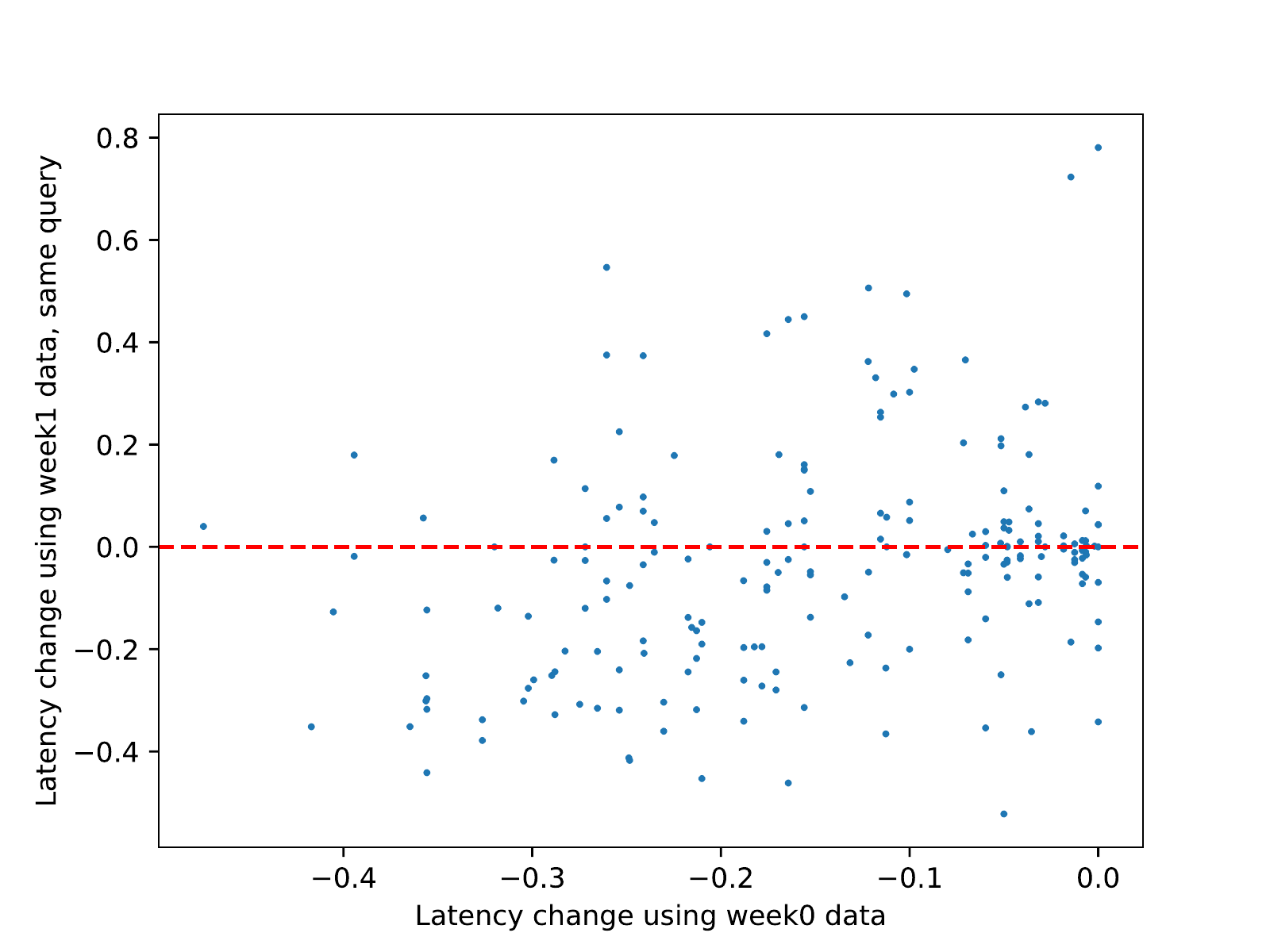}
\vspace{-6ex}
\caption{Recurring job stability: latency improvements over recurring jobs for week0 cannot always be repeated in week1.}
\label{fig:recurring_latency}
\vspace{-3ex}
\end{figure}

\paragraph{Which runtime metrics are best suited for measuring success?\\\\}

When we first tested \system in pre-production over recurring jobs, we found that even if we can find savings in latency (or PNhours) by A/B testing against the default configuration, when we run the same recurring job again on a different week, the savings cannot always be repeated, as shown in Figure~\ref{fig:recurring_latency}. For example, for a job that we found in week0 with 25\% reduction in latency ($x=-0.25$), the same recurring jobs running in week1 does not have any latency reduction, and in fact the latency became larger. Overall, we found that more than 40\% of the jobs will regress when re-run one week apart.

\begin{figure}[t]
\includegraphics[width=\linewidth]{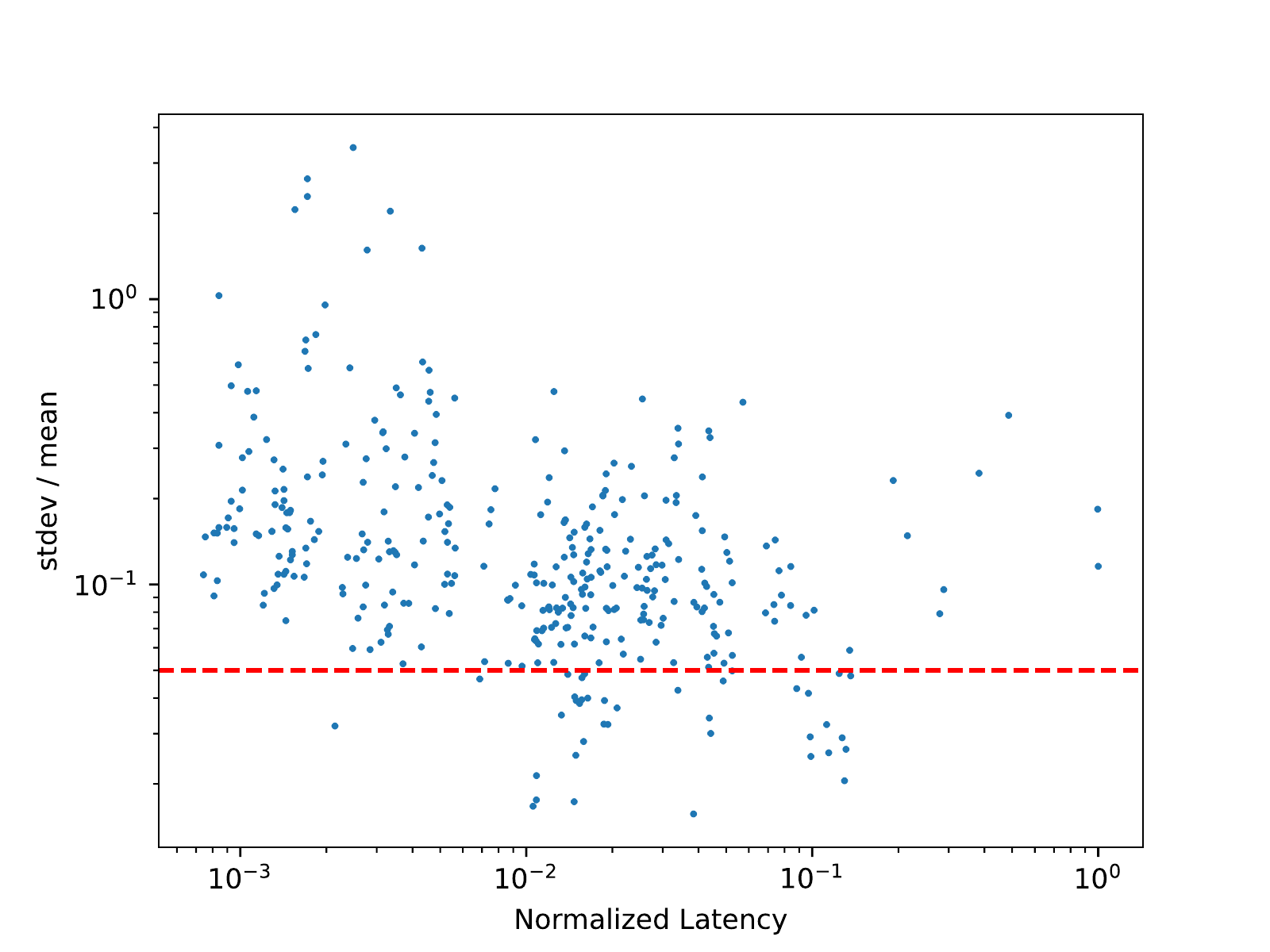}
\vspace{-6ex}
 \caption{Variance of latency plotted over jobs with different execution time (normalized). Red line is 5\% variance. The majority of jobs have high-variance.}
\label{fig:latency_variance}
\vspace{-6ex}
\end{figure}

% \stitle{Variance matter.}
After further investigation we found that the above result is due to the variance inherent in the large-scale distributed data processing systems like SCOPE. Figure~\ref{fig:latency_variance} shows an A/A test for the same jobs of Figure~\ref{fig:recurring_latency}, where we run each of them 10 times using the default rule configurations. We plot the job variance over the normalized job execution time. The red line marks 5\% variance that is typically expected by the product teams. As we can see from the Figure, more than 90\% of the jobs have more than 5\% variance in latency, with few jobs having over 100\% variance.
This variance creates problems for both learning and evaluation. For learning, datasets with such high variance contains too much noise. For evaluation, we cannot rely on a single A/B test to determine the performance improvements, and we cannot afford to always run it multiple times, and then take the mean for example.
While this variance is intrinsic in distributed processing over a cluster of machines, it is exacerbated by how SCOPE manages resources ~\cite{boutin2014apollo}.

\begin{comment}
An additional source of variability is the change in input data (both in size and distribution) across different run-days for recurring jobs.
\mi{But we should address this in the pipeline I feel.}
\zwd{We kind of capture this using data features such as estimated cardinality or total containers, but I agree this needs to be investigated further.}\aj{I agree, changes in inputs could be factored in by using them as features. We could simply skip this last sentence.} \paul{For the paper as-is, I support removing this last sentence, it raises more questions without answers.  However ... this is interesting we should be trying to predict data read and data written from the cardinality information we have about the inputs, so that observed changes in PN hours can be classified as "expected" or "surprising".  Once we switch from trying to ``gate regressions from getting in'' to ``optimistically promoting changes and detecting regressions from future runtime information`` this idea of expected vs. unexpected PN hour change will be relevant.}
\end{comment}

\begin{figure}[t]
\includegraphics[width=\linewidth]{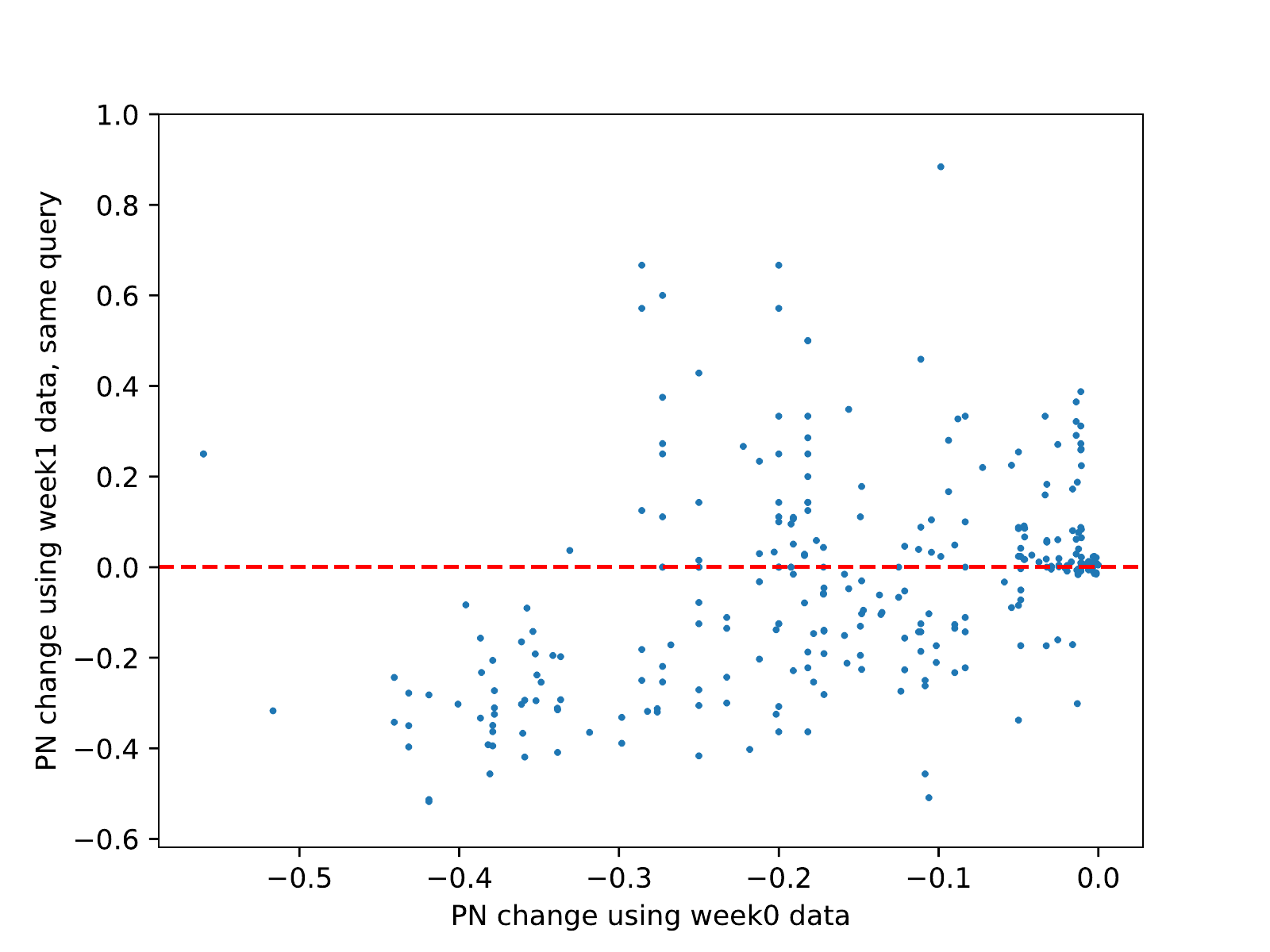}
\vspace{-6ex}
\caption{Recurring job stability: savings in PNhours over recurring jobs for week0 cannot always be repeated in week1.}
\label{fig:recurring_pn}
\vspace{-3ex}
\end{figure}

\begin{figure}[t]
\vspace{1.5ex}
\includegraphics[trim={0 0 0 4ex},clip,width=\linewidth]{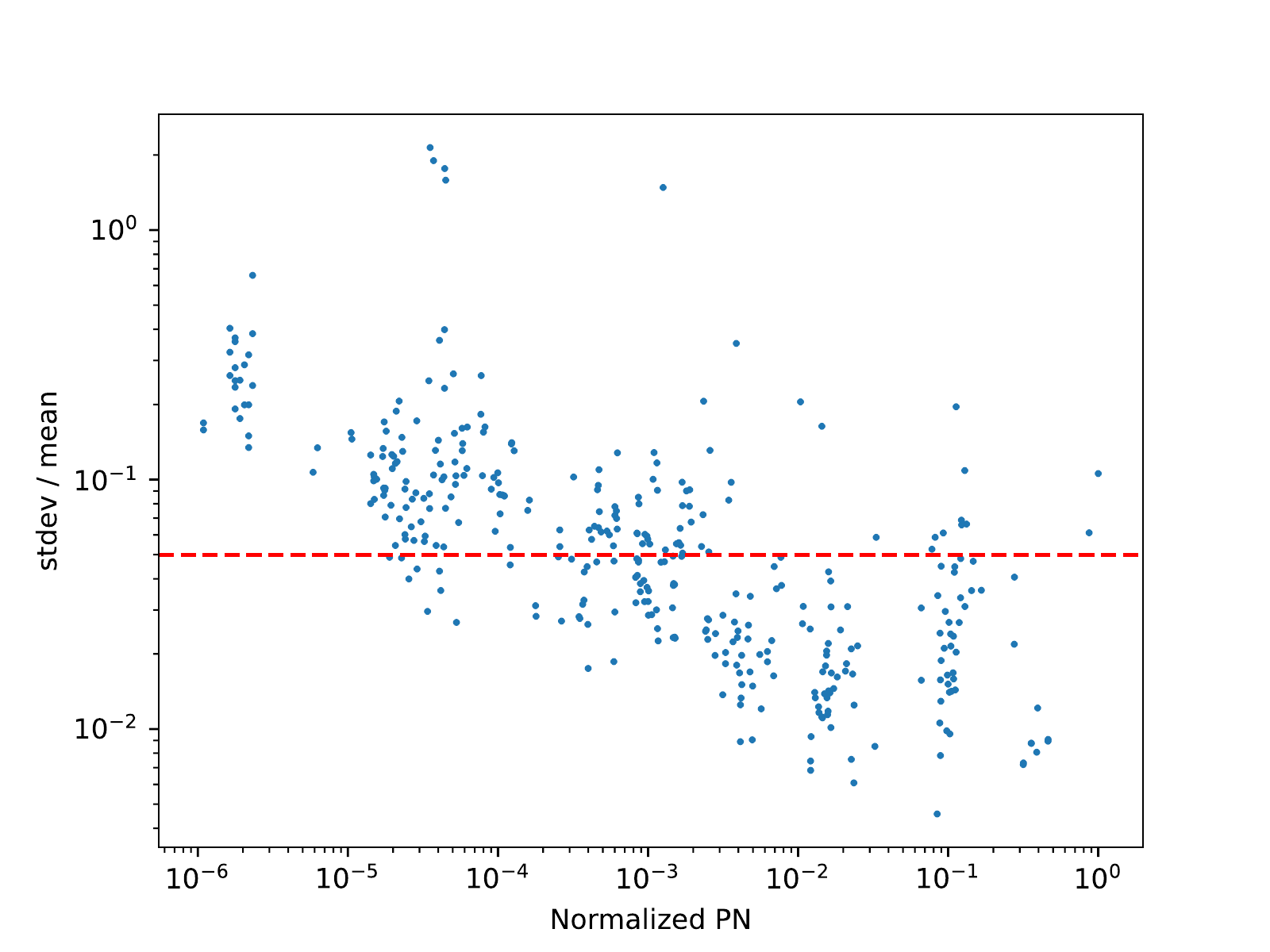}
\vspace{-6ex}
\caption{Variance of PNhours plotted over jobs with different normalized execution time. The red line is 5\% variance. Compared to the latency plot of Figure~\ref{fig:latency_variance}, PNhours is more stable, with more than 50\% jobs incurring <5\% variance.}
\label{fig:pn_variance}
\vspace{-2ex}
\end{figure}

A similar conclusion can be drawn for PNhours, as shown in Figures~\ref{fig:recurring_pn} and \ref{fig:pn_variance}. Simply relying on PNhours found previously in week0 will lead to more than 40\% regression. The variance here looks more contained however, with less than 50\% of jobs having a variance greater than 5\%.
Since the PNhours metric appears to have less variance, and there is no clear signal between estimated cost and job latency (as we shall explain next), in the current deployment of \system we decided to {\color{black}focus} on optimizing for PNhours.

\subsection{Estimated Cost vs. Real Performance}
\label{sec:experiments:estimates}
% Why do we have a pipeline?

\paragraph{Can we reliably use compile-time-only information to predict runtime outcomes?\\\\}  
Although our initial design did not include a validation component, the results in this section convinced us that some run-time information was necessary for reliable improvement.  Specifically, we tested the rule flips leading to lower estimated costs in an A/B testing using the Flighting Service.

\begin{figure}[t]
\includegraphics[width=\linewidth]{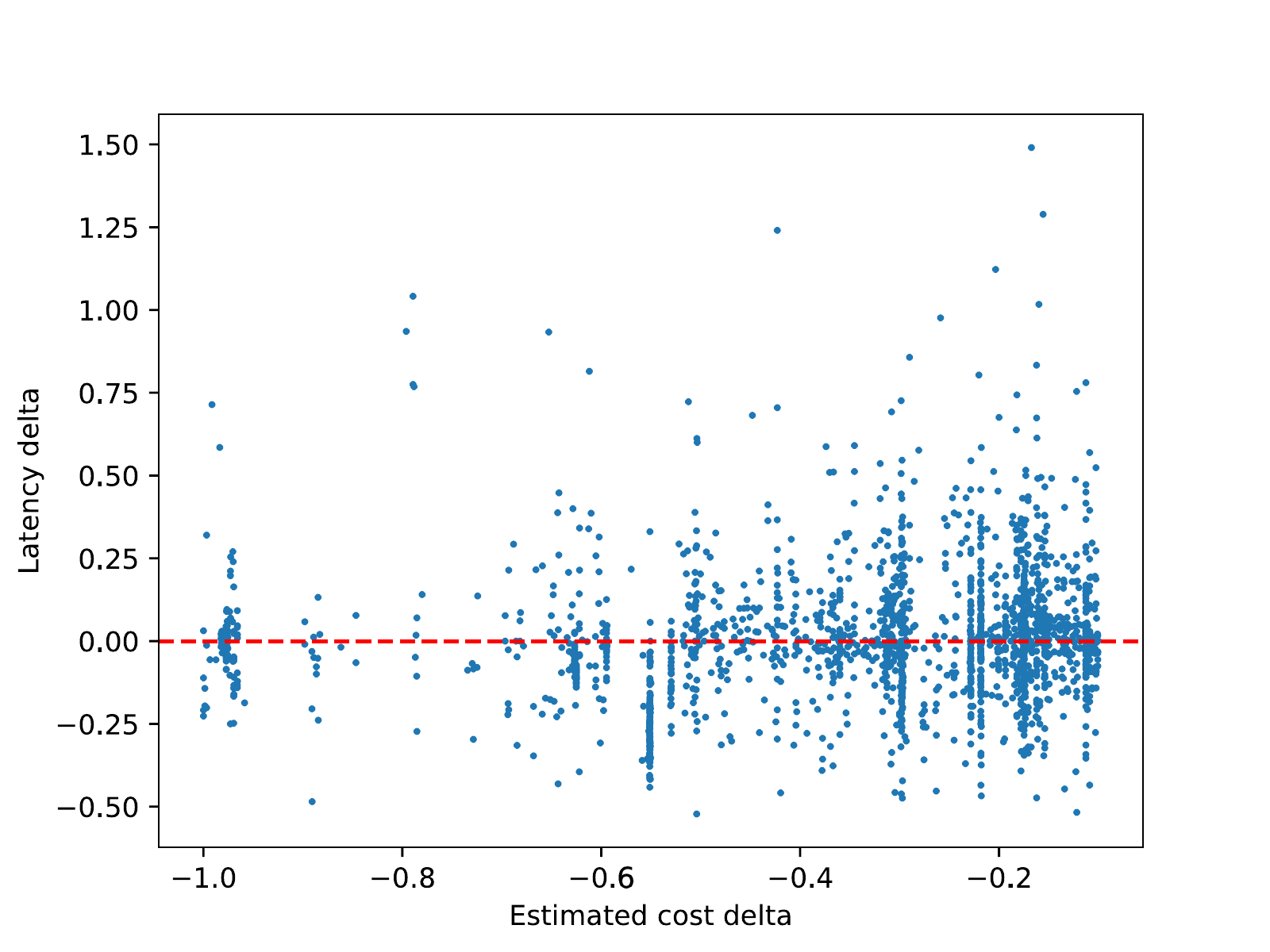}
\vspace{-6ex}
\caption{Delta (new value / old value - 1) in estimated cost versus latency. Improvement in estimated cost after recompile do not often translate in improvement in latency.}
\label{fig:cost_vs_latency}
\vspace{-3ex}
\end{figure}

Figure~\ref{fig:cost_vs_latency} plots the estimated cost delta (relative difference between the estimated cost output of the recompilation with the new hints minus the original estimated cost output of the default rule configuration), and the delta of the latency of 950 jobs run over 5 days on one of the SCOPE clusters. 
Our expectation was that as the estimated cost delta decreases (i.e., the cost decreases), the latency delta would decrease as well. However as we can see from the figure, there is no real correlation between improvements in the estimated costs and latency. For example, the left-side of the figure contains the jobs with large improvements on the estimated costs, but over 40\% of them show actually performance regression.

Nevertheless, estimated cost information are still important for the success of the approach. For instance, we ran an experiment where we disabled any use of the estimated cost coming from the SCOPE optimizer. 
Specifically, in the Recommendation step we disabled all filters based on
estimated costs, and we randomly flip one rule in the span,
instead of using the CB model. What
we found is that, after three days, \system was not able to complete
flighting (a task that usually is completed within half a day). This is
because, without using estimated costs to heuristically (1) pick
rules, and (2) filter which jobs to flight, plans leading to job latencies orders of magnitude worse than the baseline can be introduced into the pipeline, eventually making fighting impractical.

\subsection{Validation Model Performance}
\label{sec:experiments:validate}
% merge with variance section

\paragraph{Can we reliably predict future runtime outcomes from a single runtime outcome?\\\\}

As discussed in Section~\ref{sec:operationalization:validate}, we use a Validation step to verify that the recommended rule flip can indeed improve PNhours, without causing significant regression on other metrics. Since there is still some variance in the PNhours metric, we cannot trust a single run of the A/B testing for recurring jobs (Figure~\ref{fig:recurring_pn}), and, instead, we have to build a model to predict future PNhours delta.

\begin{figure}[t]
\includegraphics[width=\linewidth]{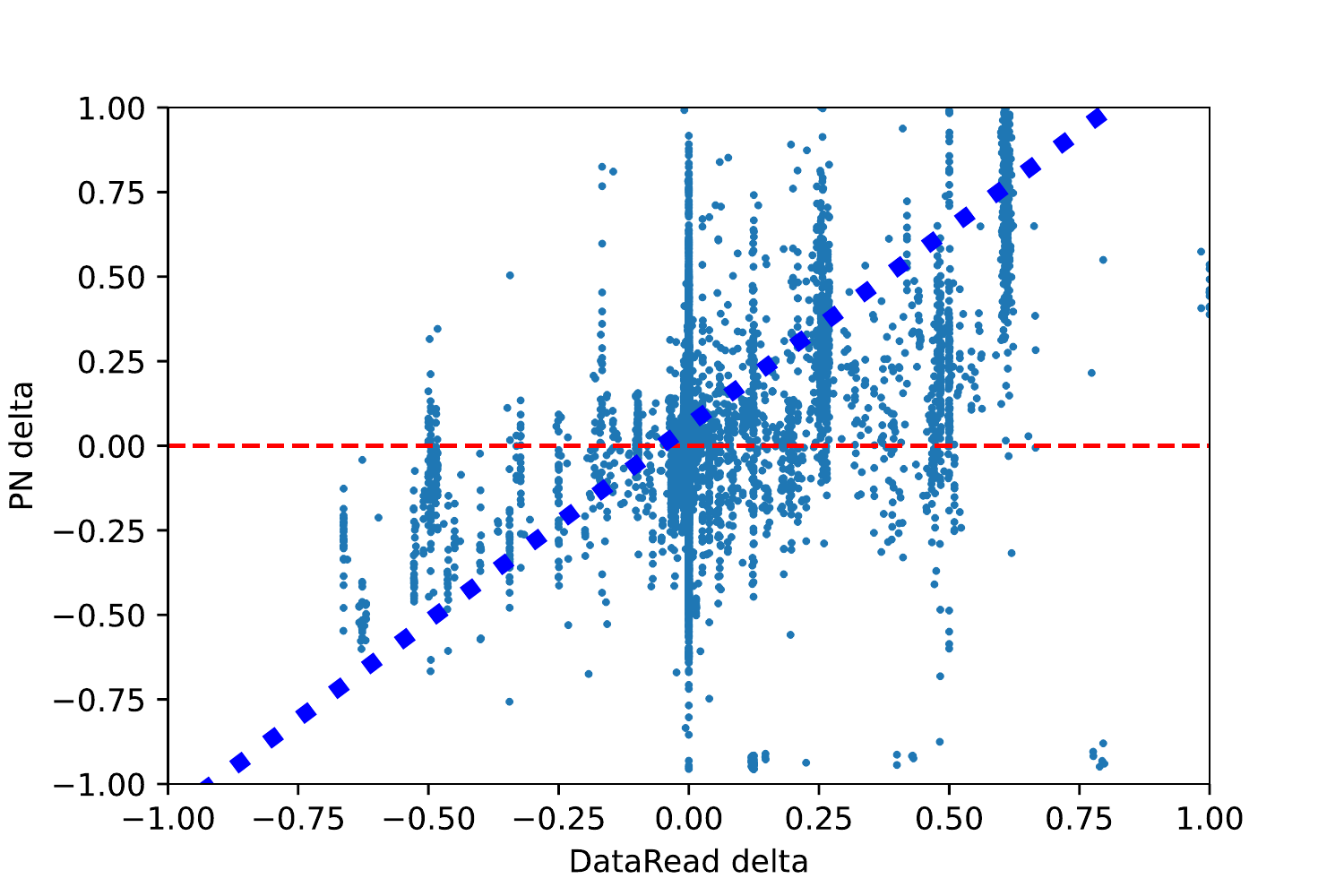}
\vspace{-6ex}
\caption{Correlation between DataRead and PNHours. Blue dotted line highlights the trend.}
\label{fig:dataread_vs_pn}
\vspace{-1ex}
\end{figure}

\begin{figure}[t]
\includegraphics[width=\linewidth]{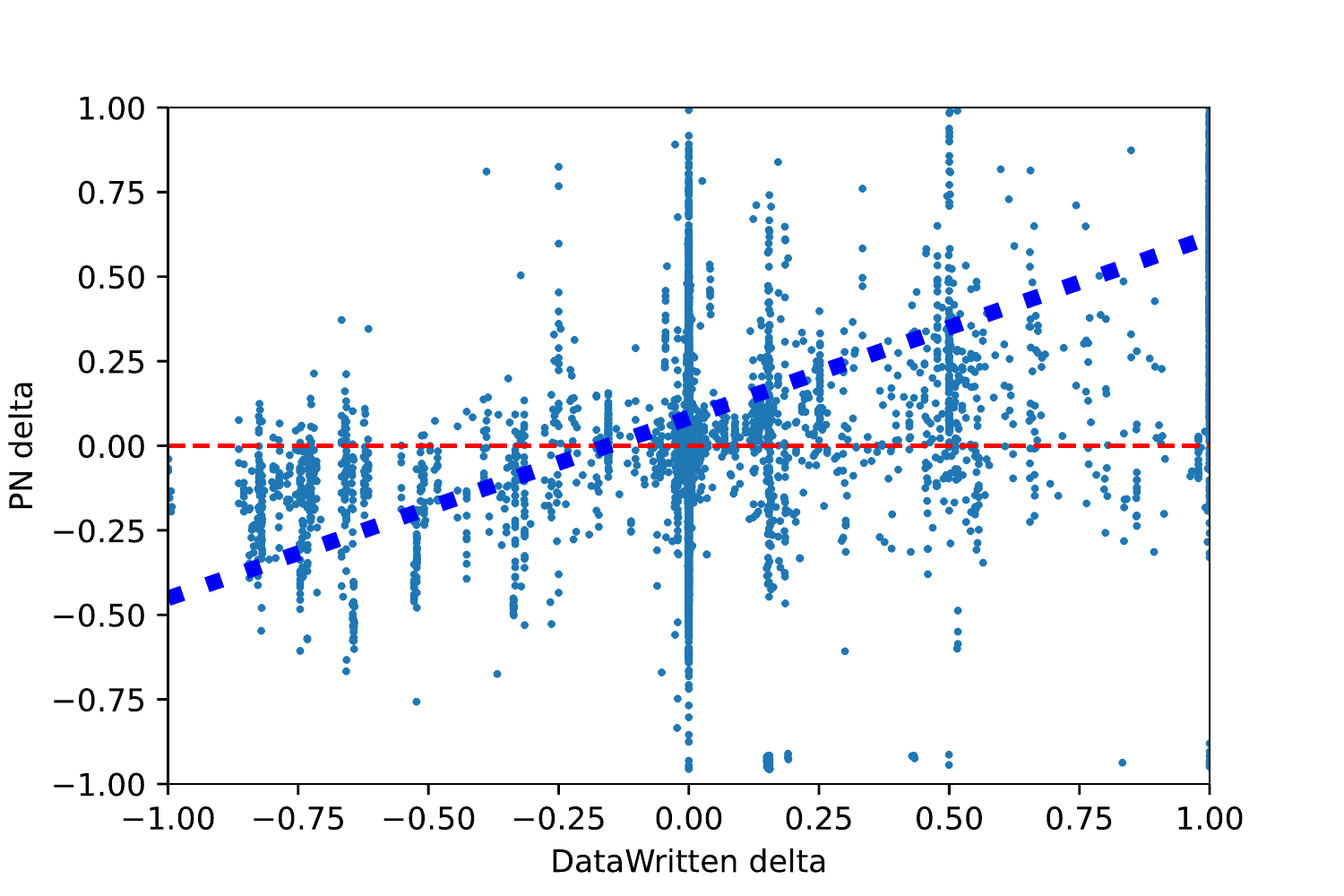}
\vspace{-6ex}
\caption{Correlation between DataWritten and PNHours. Blue dotted line highlights the trend.}
\label{fig:datawritten_vs_pn}
\vspace{-5ex}
\end{figure}

Through experimentation we found that in addition to the PNhours metric itself,  DataRead and DataWritten deltas are good indicators of how PNhours delta will change.
{\color{black}Figure}~\ref{fig:dataread_vs_pn} shows that there is a correlation between DataRead delta and PNhours delta in the historical data we gathered. %over a period of time as described in Section~\ref{sec:operationalization:validate}.
{\color{black}The dotted blue line demonstrates the trend by a one-dimensional polynomial fit.}
This result suggests that if we have seen less data read in the A/B testing, then it is likely that the PNhours will also be reduced. Similar observations can be found in Figure~\ref{fig:datawritten_vs_pn} for DataWritten. These results {\color{black}corroborate} the intuition in Section~\ref{sec:operationalization:validate} that by reducing I/O during job executions, we can reduce the PNhours and achieve better performance.

\begin{figure}[t]
\includegraphics[width=\linewidth]{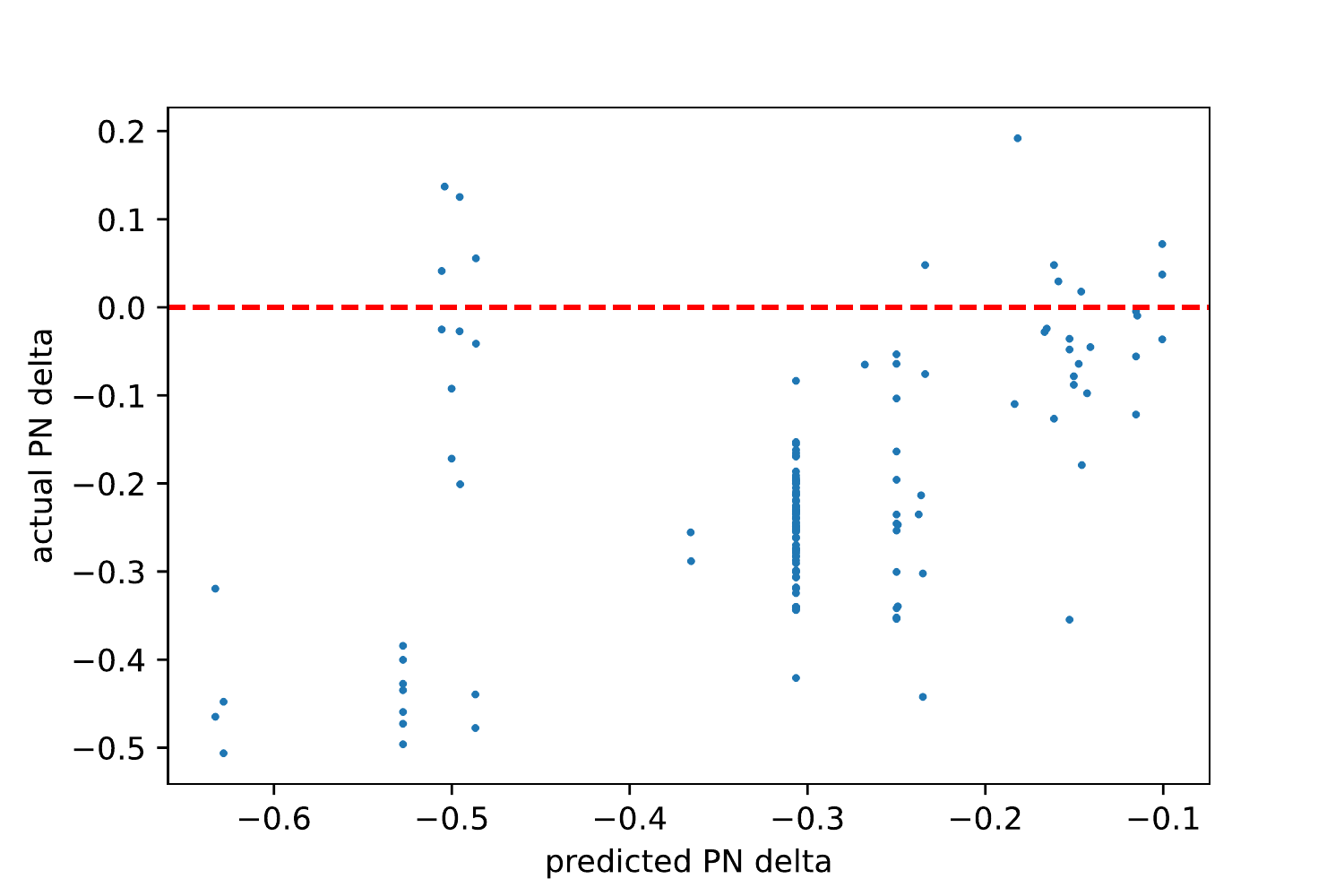}
\vspace{-6ex}
\caption{Predicted PNhours delta vs. Actual PNhours delta.}
\label{fig:predicted_vs_actual}
\vspace{-3ex}
\end{figure}

Using the Validation model described in Section~\ref{sec:operationalization:validate}, we can compute the predicted delta of PN hours. Given a certain threshold (e.g., $delta < -0.1$), we select the jobs with predicted PNhours delta qualified for this threshold, and run them in A/B testing to validate our prediction and check the results on other performance metrics. To compensate for the variance in PNhours metric, we usually use a threshold smaller than zero, so that we can avoid some regressions introduced by the variability in the cluster, as discussed in Section~\ref{sec:experiments:variance}. Figure~\ref{fig:predicted_vs_actual} demonstrates the accuracy of the Validation model (trained on historical data) on ~150 jobs in one day of the test data. For the jobs we predicted with PNhours delta smaller than -0.1, 85\% of them have their actual PNhours deltas smaller than -0.1, and 91\% of the jobs have their actual PNhours deltas smaller than 0.0 as indicated by the red line.

\subsection{Aggregate Performance}
\label{sec:experiments:flight}

\paragraph{Do we improve over default query plans in aggregate?\\\\} As introduced in Section~\ref{sec:background}, metrics of interests in our experiments include PNhours, latency, and vertices count. As just discussed in the Section~\ref{sec:experiments:variance}, in \system our primary focus is optimizing PNhours, which measures the execution cost of a SCOPE job in terms of total resources. But our goal is also to avoid significant regressions on job latency and vertices count as well.

% cosmos09, with/without recommended rule flags generated in STP
Table~\ref{tab:flighting_results} presents the aggregate pre-production results for one SCOPE workload. This workload contains 70 jobs on a single day matching the hints generated by the \system pipeline. Typically, we expect that around 5\% of  the unique jobs of the {\color{black} workloads with \system enabled} can find matches in the hints recommendation.
In aggregate, comparing to the default plans generated by the SCOPE optimizer, on this particular workload we observed 14.3\% savings in total PNhours, 8.9\% savings in total job latency, and 52.8\% savings in vertices count, after using the recommended hints for these jobs. Note that we set the threshold of predicted PN delta to -0.1, and the PNhours reductions is roughly in this range.

\begin{table}[h]
\centering
\caption{Pre-production results. 
\% of reduction for \system compared to the default rules configuration.}%\mi{To normalize.}\zwd{Not sure how beyond only saying the \%}}
\vspace{-3ex}
\label{tab:flighting_results}
\begin{tabular}{|l|l|} 
 \hline
 Metric & \%Reduction \\
 \hline\hline
 %PN hours & 46462 & 39821 & -14.3\% \\
 PNhours & -14.3\% \\
 \hline
 Latency & -8.9\% \\
 %Latency & 7158115 & 6517716 & -8.9\% \\
 \hline
 %Vertices & 20067 & 9481 & -52.8\% \\
 Vertices & -52.8\% \\
 \hline
\end{tabular}
\vspace{-3ex}
\end{table}

\subsection{Distribution of Performance Metrics}
\label{sec:experiments:distribution}

\paragraph{What is the impact on query performance in the worst case?\\\\}

\begin{figure}[t]
\vspace{-4ex}
\includegraphics[width=\linewidth]{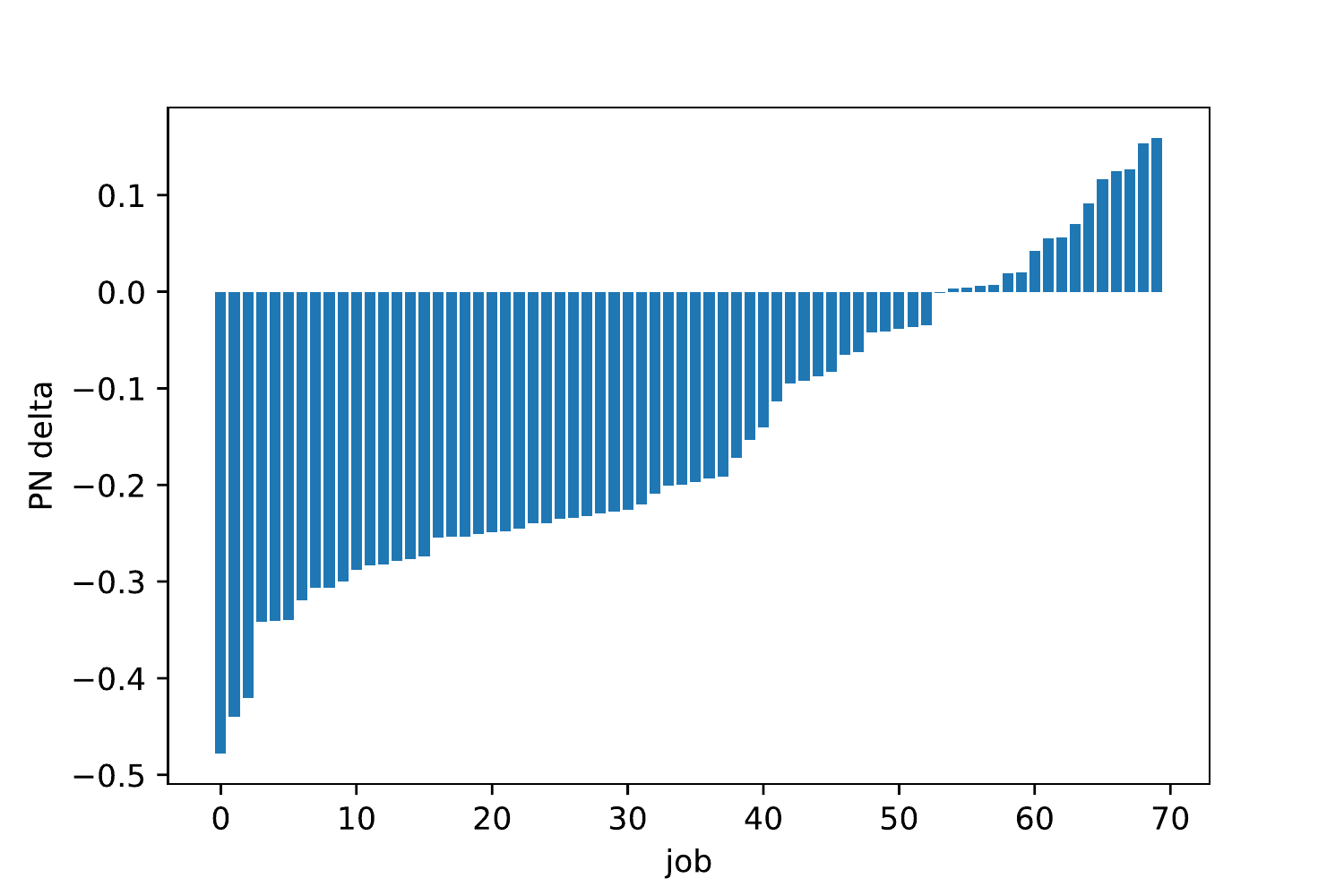}
\vspace{-6ex}
\caption{Pre-production results drill down: PNhours delta. In the best case, \system improves PNhours over the default rule configuration by approximately 50\%. In the worst case PNhours increases by ~15\%.}
\label{fig:pn_delta}
\vspace{-1.5ex}
\end{figure}

\begin{figure}[t]
\vspace{-4ex}
\includegraphics[width=\linewidth]{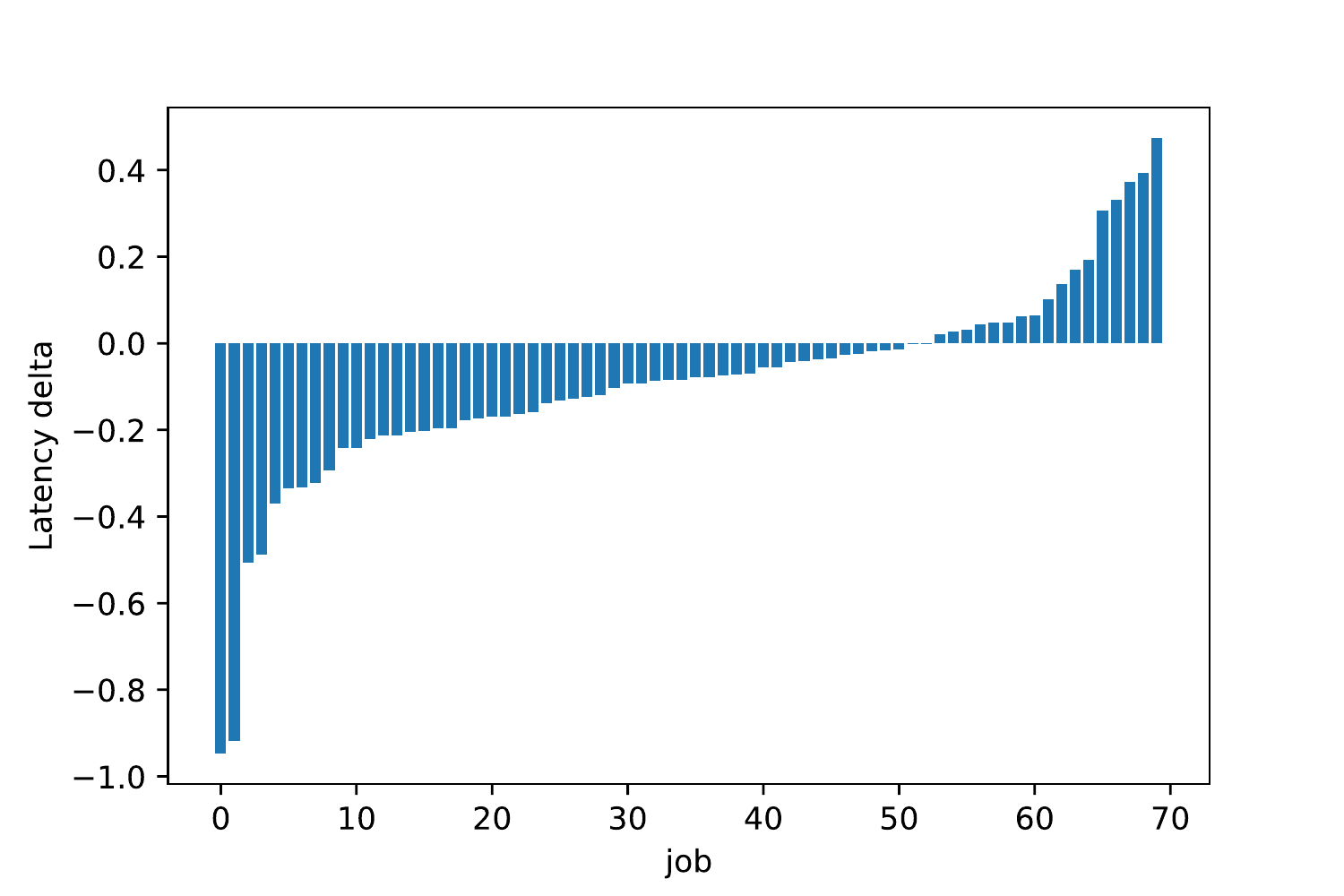}
\vspace{-6ex}
\caption{Pre-production results drill down: latency delta. In the best case \system is able to improve latency by ~90\%. In the worst case, \system introduces a regression of about 45\%. The larger regression compared to PNhours (Figure~\ref{fig:pn_delta}) is because \system is tuned over PNhours.}
\label{fig:latency_delta}
\vspace{-7ex}
\end{figure}

\begin{figure}[t]
\includegraphics[width=\linewidth]{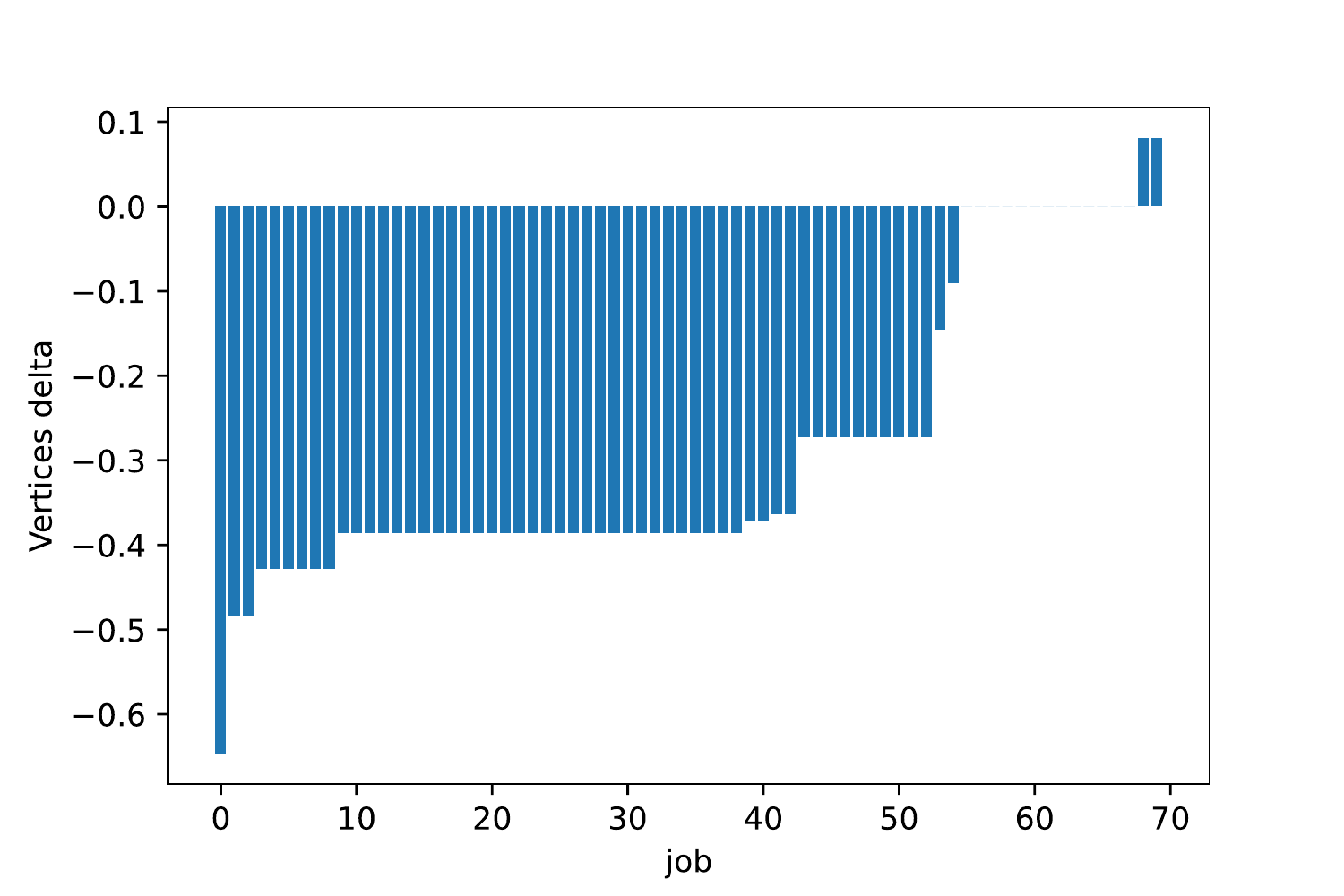}
\vspace{-6ex}
\caption{Pre-production results drill down: vertices delta. In this case \system introduces a regression of 10\% in the worst case, while in the best case it can improve vertices utilization by more than 60\%.}
\label{fig:vertices_delta}
\vspace{-3ex}
\end{figure}

In this section we drill down on the aggregated results presented in the previous Section.
Figures~\ref{fig:pn_delta}, \ref{fig:latency_delta}, and \ref{fig:vertices_delta} present the performance changes for the 70 jobs of Section~\ref{sec:experiments:flight} on three metrics (PNhours, latency, and vertices, respectively) compared with the default query plan. Note that in each figure, the jobs are ordered by the change in that specific metric, and a delta greater than 0.0 indicates a regression.

In Figure~\ref{fig:pn_delta}, we see that over 80\% of the jobs have smaller PNhours, indicating savings in resource usage. In the best case on the left part of the figure, the PNhours of a job is reduced by almost 50\%. 
Conversely, in the worst case, the PNhours of a job is increased by at most ~15\%---a considerable improvement over the 100\% worst case regression of our previous work that uses randomly chosen rule flips without any validation~\cite{steerqo}.%.\paul{Is there a quick number we can put from the previous work for comparison?  Otherwise it's just ``better and go check the reference if you care about how much ...''}

Regarding job latency, in Figure~\ref{fig:latency_delta} we also observe that about 80\% of jobs have reduced their latency, by a maximum of over 90\%.
We also see that 20\% of jobs have latency increased, but compared with the results in Figure~\ref{fig:cost_vs_latency} (using estimated cost alone) or Figure~\ref{fig:recurring_pn} (using recurring jobs without validation), the regression rate is reduced from more than 40\% to 20\%. However, since in the Validation step we primarily focused on optimizing for PNhours, the regression on latency is less contained.

Finally, for vertices count (Figure~\ref{fig:vertices_delta}), we see that only two jobs used 10\% more vertices. This effect appears to be explainable by our approach of optimizing PNhours and reducing the data that is read and written during job execution, since the I/O reduction might be a natural result of fewer vertices used for job execution which consequentially requires less data transmission.

So given the small fraction of regressions and the contained impact of the regressed jobs, we consider that \system is working well on production workloads and overall we can reduce the PNhours, job latency and vertices for the recurring jobs matching our recommendations.

\subsection{Biased Randomization}
\label{sec:experiments:recompile}

\paragraph{Is our biased search using contextual bandit more effective than a uniformly-at-random baseline?\\\\}
Conceivably, the Validation step of the \system pipeline might be strong enough to ensure aggregate improvement, even if the biased randomization in the Recommendation component provides no improvement. Therefore, we evaluated the effectiveness of the rule Recommendation step (Section~\ref{sec:operationalization:recommend}) using {\color{black}CB} against a baseline that chooses a rule uniformly at random in the action set to flip.
For this experiment, we check how often these two approaches can find better query plans in terms of optimizer estimated costs, which we use as the reward to train the reinforcement learning policy.

For the workload tier we are currently running \system on, from one single day
after feature generation (Section~\ref{sec:operationalization:feature}) around 66\% of the jobs have non-empty job span (i.e. the action set). For these jobs, the random baseline flips one rule out of the span at uniform random, while our approach chooses one rule with the guidance of the contextual bandit policy. Note that for the contextual bandit policy, the selected rule flip does not necessarily lead to a better plan than the default rule configuration.  Thus we always recompile with the {\color{black}CB's} selected rule flip and short-circuit subsequent processing if there is no estimated cost improvement.

For the same set of jobs in the workload, we recompile with the rule flips to check whether the estimated costs of the jobs become lower, higher, or remain the same compared to the default rule configurations. Table~\ref{tab:random_vs_CB} shows the number of jobs in each category of the cost changes. We observe that using {\color{black}CB} increases the number of jobs with lower costs by 3$\times$, and reduces the number of jobs with higher costs by almost 2$\times$. It also leads to fewer recompilation failures. Overall, the total estimated cost of this particular workload is reduced from $1.7e11$ using random rule flips to $1.0e09$ using the {\color{black}CB} rule flips---an improvement of over 100$\times$.

\begin{table}[t]
\centering
\caption{Comparison between random and {\color{black}CB} rule flips.}
\vspace{-3ex}
\label{tab:random_vs_CB}
\begin{tabular}{|l|l|l|l|l|}
 \hline
 Number of jobs & Random & Random \% & CB & CB\% \\
 \hline\hline
 Lower cost & 377 & 10.6\% & 1226 & 34.5\% \\ 
 \hline
 Equal cost & 1257 & 35.4\% & 1141 & 32.1\% \\
 \hline
 Higher cost & 1280 & 36.0\% & 693 & 19.5\% \\
 \hline
 Recompile failures & 638 & 18.0\% & 492 & 13.9\% \\
 \hline
\end{tabular}
\vspace{-3ex}
\end{table}

% % TODO: pie chart
% \begin{figure}[h]
% \includegraphics[width=0.4\textwidth]{figures/random_vs_RL.png}
% \caption{Comparison of random and CB rule flips}
% \label{fig:random_vs_CB}
% % \vspace{-3ex}
% \end{figure}

% TODO: Need some way to better quantify the accuracy.

% \subsection{Feature Importance}

% feature importance, e.g. loss during CFE with and without certain features.

% \subsection{Production Results}
% \label{sec:experiments:production}

% efficiency: how long each step takes
% - mostly flighting (and partly because we don't use full capacity), many hours
% - then span generation, 1st iteration 2 hours, then ~30min for a few more iterations
% - then recompilation during training, 2 hours
% - training the CB model is fast (and happens at personalizer separately)

% production results: relative percentage over recurring jobs, and delta before/after applying the STP.

% personalizer backend results.

% \mi{This is what I have in mind to show (at a minimum):
% \begin{itemize}
%     \item Pre-production results (basically what we have in the presentation right now)
%     \item Production results over tier 4
%     \item Feature importance for each model
% \end{itemize}
% In each case say how many jobs we get as input, and how many we consider for each stage in the pipeline.}

\section{Lessons Learned}
\label{sec:lessons}

\stitle{From research to practice: trade-offs.}
Applying academic research ideas into production always involves trade-off,  and going from the BAO's ideas to \system was no exception.
For example, already in~\cite{steerqo} we realized that generating new rules configurations using runtime information at SCOPE scale requires a large amount of resources. Not just because SCOPE workloads are orders of magnitude larger than the one considered in ~\cite{Marcus2020BaoLT}, but also because SCOPE workloads are more complex, and only focusing on few rules (as in BAO) would not work for all use-cases. 
{\color{black}In our initial investigation, in fact, we found that there is no subset of rules that captures all the benefits, whereas all the rules need to be considered to properly optimize SCOPE workloads.}
In \system we therefore decided to focus on all the 256 rules available in SCOPE, and to learn over estimated costs (which can be quickly returned by the SCOPE optimizer) rather than actual runtime metrics.

\stitle{Simplicity first.}
Similarly, we found that suggesting arbitrary complex rule configurations such as in ~\cite{steerqo} was not ideal for our product partners because in case of regression it would be almost impossible to debug. 
We therefore decided to stage the problem with only one rule difference from the default rule configuration in the first version.
This allowed us to build trust with the product team where this feature works, and also explain better which rules are really moving the needle, something the team could further investigate and corroborate based on their domain knowledge.

\stitle{Do not reinvent the wheel.}
In the first prototype of \system, we used Vowpal Wabbit~\cite{vw} for the contextual bandit model. However, as we hardened the system, we realized that maintaining the state over pipeline runs in a reliable way is non-trivial, especially if, for explainability, we want to trace how the model evolves and learns over time.
Instead of building our own infrastructure, we decided to integrate \system with Azure Personalizer~\cite{agarwal2017making}, which manages the log and state, as well as provides all the enterprise-grade features we needed for our scenario.

\stitle{Regressions.}
We learned  that performance regressions are important to catch upfront. At SCOPE scale, if even a 1\% of the jobs introduce regression, the number of customer incidents could easily overwhelm the product team~\cite{vldb-tutorial-2021}. 
At the same time, we learned that SCOPE clusters have non-negligible variability in performance, making it practically impossible to effectively catching regression without extensive A/A and A/B testing.
Our trade-off here is to try to catch regression in a best effort way, with the limited budged allowed for flighting, and by using a machine learning model to validate the output of flighting before uploading the actions for online production consumption.
In general, live online experimentation is expensive, difficult, or even not allowed to be done in SCOPE-like production systems that run critical customer workloads. 
We use counter-factual evaluations where we can rely on past telemetry offline to improve learning parameters and to tune the model.

\stitle{The surprising effectiveness of span features.} 
Interestingly, we found that complex featurizations of query plans, as suggested in previous works (e.g.,~\cite{marcus2019neo,Marcus2020BaoLT}) were mostly ineffective, whereas %a simple feature such as the job span is already a good proxy for query plans in our settings.
the use of the complete job span as context features was critical to our success. 
In particular second and third order coocurrence indicator features over the job span.  Intuitively, the complete job span concisely represents the query plan by identifying the set of controls which affect the optimizer output \emph{on this instance}.   This strategy is plausibly reusable in other scenarios where a learning system is attempting to steer the result of an optimization subroutine in an instance-dependent fashion. 
%\zwd{Maybe this is a detail we can say in sec 3 or 4, or we could say some thing about featurization and counter-factual evaluations offline.}

\section{Related Work}
\label{s:related}

%\rev{Alekh, Matteo, Wangda and Paul}

\stitle{Steering query optimizers.}
%Bao~\cite{Marcus2020BaoLT} was the early work that leveraged PostgreSQL query hints to 
BAO~\cite{Marcus2020BaoLT} generates $48$ rule configurations in PostgreSQL 
%with each rule configuration essentially instantiating a simpler version of the PostgreSQL query optimizer. Each simple optimizer disables a subset of the rules 
that affect the optimizer behaviour for choosing scan operators, join operators, and join orders. BAO treats each rule configuration as an arm in a multi-armed bandit problem and given a new query, it learns to choose one of the $48$ arms. BAO models a reinforcement learning problem in which it sees a sequence of queries, and over time learns to make the correct decisions. BAO was evaluated on a custom dataset with queries ranging from a few to several hundred seconds. 
%Bao did not rely on the PostgreSQL cost model. 
Also, a core component of BAO is a tree convolutional neural network which learns a cost model for tree structured PostgreSQL query plans by executing the plans.

In our recent work~\cite{steerqo}, we extended the steering optimizer research by considering 
how an approach similar to BAO can be ported over real-world workloads in industry strength cloud data processing system, SCOPE~\cite{chaiken2008scope}. 
We show that indeed it is possible to improve SCOPE plans by steering its query optimizer, although this requires substantial experimentation and resources since SCOPE optimizer rule set is quite large. Another problem that we faced is that, while plans can be improved, a large portion of the explored plans actually introduce regressions. In this work we address both these concerns, and show, through production results, that is possible to steer SCOPE optimizer towards better plans, without necessarily introduce too much regressions.

\vspace{1ex}
\stitle{Learning-based optimizations in SCOPE.}
%\mi{Taken from~\cite{steerqo}, needs a refresh}
%SCOPE is large-scale data processing platform using by the various business units at Microsoft. We refer the readers to early papers on SCOPE for its  overall design~\cite{scope}, and to a recent paper recapping its journey over the last two decades~\cite{cosmoshistory}. 
Apart from the scale and complexity, one of the other major challenges in SCOPE is efficiency~\cite{patel2019big}.
As a result, the Peregrine infrastructure has been built over the recent years to introduce workload optimizations in SCOPE to optimize multiple jobs and reduce the overall costs~\cite{jindal2019peregrine}. 
Subsequently, several efforts have built on top of Peregrine to introduce learned components in SCOPE. Examples include learned models to automatically choose the number of concurrent containers a job should use~\cite{autotoken, tasq, bag2020towards}, learned cardinality estimation tailored to the past SCOPE workloads~\cite{wu2018towards}, 
a learning-based approach for cost models~\cite{siddiqui2020cost}, and even learning-based checkpointing~\cite{phoebe}.

\vspace{1ex}
\stitle{Instance-optimized database components.}
%\mi{something on all the work on learning. Alekh please add.}
Several recent works focus on improving database components~\cite{mlos,pavlo17,vanaken17,DBLP:conf/cidr/KraskaABCKLMMN19,DBLP:conf/sigmod/KraskaBCDP18} or generating efficient query plans~\cite{rejoin,marcus2019neo,berkeley_drl} by exploiting learning over actual runtimes or the output of an optimizer’s cost model.
%Conceptually, this is similar to the past work exploring the space of query plans, such as Picasso~\cite{haritsa2010picasso}, or designing robust cost estimates for choosing query plans~\cite{wolf2018robustness}.
While instance optimization is very appealing for modern cloud customers, there are concerns about its end to end integration, impact on actual performance, and also potential performance regressions~\cite{perfguard, ai-meets-ai}. 

\vspace{1ex}
\stitle{Contextual Bandit, Reinforcement Learning and their applications.}
Applications and systems constantly face decisions that require picking from a set of actions based on contextual information. Reinforcement-based learning algorithms such as contextual bandits~\cite{langford2007epoch} can be very effective in these settings. The Azure Personalizer service~\cite{agarwal2017making} implements several contextual bandits algorithms and is deployed to many content recommendation applications such as MSN news recommendation and XBox website personalization. For systems, deployed instances of contextual bandit approaches include Azure Compute that wants to contextually optimize the waiting time with non-responsive virtual machines in a cloud service, and tuning parameters of components in Skype~\cite{gupchup2020resonance}.
\section{Limitations and Future Work}
\label{sec:future}

In its current version, \system has several limitations that we plan to improve as we build deployment experience and confidence in the pipeline.
The first limitation is that \system currently suggests only one single rule flip per job. While the simplicity of this solution helps with debugging and with building trust with the product partners, it limits how many plans can be improved.  In future work we will propose multiple rule flips, e.g., by utilizing techniques from combinatorial contextual bandits or short-horizon episodic reinforcement learning~\cite{sutton2018reinforcement}.

Another limitation is the inefficiency of the Validation stage of the pipeline, which requires comparing the runtime metrics of the proposed rule change to the default configuration.  Even for recurrent queries, where this additional query execution cost can be amortized over the lifetime of the query, the overhead limits the absolute number of proposed rule changes we can accept per day due to capacity constraints.  In future work we will attempt to optimistically accept proposed query plans and detect regressions from subsequent runtime metrics.

Other limitations include: (1)~we are currently using heuristics to pick which jobs to flight, and (2)~each pipeline is run on a given date independently from the others.
We are exploring how to make \system stateful such that jobs already explored on a previous date can be skipped, when unmodified, in successive dates.
In the long term, we are planning to make flighting part of the model, such that we can chose which job to flight based on what part of the space the model thinks is better to explore.
Eventually, we would like to go beyond recurrent jobs and provide  hints for any SCOPE jobs. This will require a tighter integration between the SCOPE optimizer and Azure Personalizer, in an online learning setting assisted by an offline pipeline for exploring new configurations.

Finally, \system is just one of many learned components that are currently being explored in production to improve SCOPE performance. It is not clear how these learned components synergize with each other, and this is an exciting area requiring further exploration.
For instance, the quality of query plans is directly related to the quality of the estimated costs and cardinalities, but these estimates can be off by a large margin in practice. 
\system avoids this pitfall by learning how to directly steer the query optimizer towards better plans. %by learning how different rule configurations improve estimated costs and runtime metrics.
Will better learned cardinalities improve the plans generated by \system, and vice-versa, in a virtuous cycle? Or as we add more learned components, will the improvements coming from each feature compound with each other?

% \mi{
% Just be honest that we have a predictor for detecting regressions, so we made progress here,but we need more work here.
% From naive experiments design to more informed one.}
\section{Conclusion}
\label{sec:conclusion}

%The idea of directly steering query optimizers towards better plans by turning rules on/off went long way since the idea was first introduced in~\cite{Marcus2020BaoLT}.
We provide an update on our journey towards making steering query optimizers a reality in production for SCOPE users.
We introduce \system to address several shortcomings of previous approaches in a novel way. 
Specifically, we use a pipeline of models to (1) generate new rule configurations 1-edit distance from the default configuration; and (2) validate A/B testing runs to catch possible regressions before production deployment.
As of November 2021, \system pipeline runs offline periodically (once a day) and is currently deployed in production over different SCOPE clusters. %for one of their workload tier.

\begin{acks}
We would like to thank Carlo Curino, John Langford, Raghu Ramakrishnan, and Siddhartha Sen for their insightful feedback, as well as GT Ni for the work during early stages of the development.
\end{acks}

% \vspace{-2ex}
% \section{Acknowledgements}\vspace{-2ex}
% We thank the anonymous reviewers and our shepherd, Chen Wenguang, for their feedback and suggestions to improve the paper. We would also like to thank Nellie Gustafsson, Gopal Vashishtha, Emma Ning, and Faith Xu for their support.

\balance

\bibliographystyle{ACM-Reference-Format}
%\bibliography{}  % sigproc.bib is the name of the Bibliography 
\bibliography{main}


\begin{thebibliography}{43}

%%% the Web address.

\ifx \showCODEN    \undefined \def \showCODEN     #1{\unskip}     \fi
\ifx \showDOI      \undefined \def \showDOI       #1{#1}\fi
\ifx \showISBNx    \undefined \def \showISBNx     #1{\unskip}     \fi
\ifx \showISBNxiii \undefined \def \showISBNxiii  #1{\unskip}     \fi
\ifx \showISSN     \undefined \def \showISSN      #1{\unskip}     \fi
\ifx \showLCCN     \undefined \def \showLCCN      #1{\unskip}     \fi
\ifx \shownote     \undefined \def \shownote      #1{#1}          \fi
\ifx \showarticletitle \undefined \def \showarticletitle #1{#1}   \fi
\ifx \showURL      \undefined \def \showURL       {\relax}        \fi
% invisible to TeX
\providecommand\bibfield[2]{#2}
\providecommand\bibinfo[2]{#2}
\providecommand\natexlab[1]{#1}
\providecommand\showeprint[2][]{arXiv:#2}

\bibitem[\protect\citeauthoryear{Agarwal, Bird, Cozowicz, Hoang, Langford, Lee,
  Li, Melamed, Oshri, Ribas, et~al\mbox{.}}{Agarwal et~al\mbox{.}}{2016}]%
        {agarwal2017making}
\bibfield{author}{\bibinfo{person}{Alekh Agarwal}, \bibinfo{person}{Sarah
  Bird}, \bibinfo{person}{Markus Cozowicz}, \bibinfo{person}{Luong Hoang},
  \bibinfo{person}{John Langford}, \bibinfo{person}{Stephen Lee},
  \bibinfo{person}{Jiaji Li}, \bibinfo{person}{Dan Melamed},
  \bibinfo{person}{Gal Oshri}, \bibinfo{person}{Oswaldo Ribas},
  {et~al\mbox{.}}} \bibinfo{year}{2016}\natexlab{}.
\newblock \showarticletitle{Making contextual decisions with low technical
  debt}.
\newblock \bibinfo{journal}{\emph{arXiv preprint arXiv:1606.03966}}
  (\bibinfo{year}{2016}).
\newblock


\bibitem[\protect\citeauthoryear{Agarwal, Hsu, Kale, Langford, Li, and
  Schapire}{Agarwal et~al\mbox{.}}{2014}]%
        {agarwal2014taming}
\bibfield{author}{\bibinfo{person}{Alekh Agarwal}, \bibinfo{person}{Daniel
  Hsu}, \bibinfo{person}{Satyen Kale}, \bibinfo{person}{John Langford},
  \bibinfo{person}{Lihong Li}, {and} \bibinfo{person}{Robert Schapire}.}
  \bibinfo{year}{2014}\natexlab{}.
\newblock \showarticletitle{Taming the monster: A fast and simple algorithm for
  contextual bandits}. In \bibinfo{booktitle}{\emph{International Conference on
  Machine Learning}}. PMLR, \bibinfo{pages}{1638--1646}.
\newblock


\bibitem[\protect\citeauthoryear{Agrawal, Chaudhuri, Kollar, Marathe,
  Narasayya, and Syamala}{Agrawal et~al\mbox{.}}{2005}]%
        {sql-server-advisor}
\bibfield{author}{\bibinfo{person}{Sanjay Agrawal}, \bibinfo{person}{Surajit
  Chaudhuri}, \bibinfo{person}{Lubor Kollar}, \bibinfo{person}{Arun Marathe},
  \bibinfo{person}{Vivek Narasayya}, {and} \bibinfo{person}{Manoj Syamala}.}
  \bibinfo{year}{2005}\natexlab{}.
\newblock \showarticletitle{Database tuning advisor for microsoft sql server
  2005}. In \bibinfo{booktitle}{\emph{Proceedings of the 2005 ACM SIGMOD
  international conference on Management of data}}. \bibinfo{pages}{930--932}.
\newblock


\bibitem[\protect\citeauthoryear{Ammerlaan, Antonius, Friedman, Hossain,
  Jindal, Orenberg, Patel, Qiao, Ramani, Rosenblatt, et~al\mbox{.}}{Ammerlaan
  et~al\mbox{.}}{2021}]%
        {perfguard}
\bibfield{author}{\bibinfo{person}{Remmelt Ammerlaan}, \bibinfo{person}{Gilbert
  Antonius}, \bibinfo{person}{Marc Friedman}, \bibinfo{person}{HM~Sajjad
  Hossain}, \bibinfo{person}{Alekh Jindal}, \bibinfo{person}{Peter Orenberg},
  \bibinfo{person}{Hiren Patel}, \bibinfo{person}{Shi Qiao},
  \bibinfo{person}{Vijay Ramani}, \bibinfo{person}{Lucas Rosenblatt},
  {et~al\mbox{.}}} \bibinfo{year}{2021}\natexlab{}.
\newblock \showarticletitle{PerfGuard: deploying ML-for-systems without
  performance regressions, almost!}
\newblock \bibinfo{journal}{\emph{Proceedings of the VLDB Endowment}}
  \bibinfo{volume}{14}, \bibinfo{number}{13} (\bibinfo{year}{2021}),
  \bibinfo{pages}{3362--3375}.
\newblock


\bibitem[\protect\citeauthoryear{Azure}{Azure}{[n.d.]}]%
        {adf}
\bibfield{author}{\bibinfo{person}{Microsoft Azure}.}
  \bibinfo{year}{[n.d.]}\natexlab{}.
\newblock \bibinfo{title}{Azure Data Factory}.
\newblock
  \bibinfo{howpublished}{\url{https://azure.microsoft.com/en-us/services/data-factory/}}.
\newblock


\bibitem[\protect\citeauthoryear{Bag, Jindal, and Patel}{Bag
  et~al\mbox{.}}{2020}]%
        {bag2020towards}
\bibfield{author}{\bibinfo{person}{Malay Bag}, \bibinfo{person}{Alekh Jindal},
  {and} \bibinfo{person}{Hiren Patel}.} \bibinfo{year}{2020}\natexlab{}.
\newblock \showarticletitle{Towards plan-aware resource allocation in
  serverless query processing}. In \bibinfo{booktitle}{\emph{12th USENIX
  Workshop on Hot Topics in Cloud Computing (HotCloud 20)}}.
\newblock


\bibitem[\protect\citeauthoryear{Boutin, Ekanayake, Lin, Shi, Zhou, Qian, Wu,
  and Zhou}{Boutin et~al\mbox{.}}{2014}]%
        {boutin2014apollo}
\bibfield{author}{\bibinfo{person}{Eric Boutin}, \bibinfo{person}{Jaliya
  Ekanayake}, \bibinfo{person}{Wei Lin}, \bibinfo{person}{Bing Shi},
  \bibinfo{person}{Jingren Zhou}, \bibinfo{person}{Zhengping Qian},
  \bibinfo{person}{Ming Wu}, {and} \bibinfo{person}{Lidong Zhou}.}
  \bibinfo{year}{2014}\natexlab{}.
\newblock \showarticletitle{Apollo: Scalable and Coordinated Scheduling for
  Cloud-Scale Computing}. In \bibinfo{booktitle}{\emph{11th USENIX Symposium on
  Operating Systems Design and Implementation (OSDI 14)}}.
  \bibinfo{pages}{285--300}.
\newblock


\bibitem[\protect\citeauthoryear{Chaiken, Jenkins, Larson, Ramsey, Shakib,
  Weaver, and Zhou}{Chaiken et~al\mbox{.}}{2008}]%
        {chaiken2008scope}
\bibfield{author}{\bibinfo{person}{Ronnie Chaiken}, \bibinfo{person}{Bob
  Jenkins}, \bibinfo{person}{Per-{\AA}ke Larson}, \bibinfo{person}{Bill
  Ramsey}, \bibinfo{person}{Darren Shakib}, \bibinfo{person}{Simon Weaver},
  {and} \bibinfo{person}{Jingren Zhou}.} \bibinfo{year}{2008}\natexlab{}.
\newblock \showarticletitle{Scope: easy and efficient parallel processing of
  massive data sets}.
\newblock \bibinfo{journal}{\emph{Proceedings of the VLDB Endowment}}
  \bibinfo{volume}{1}, \bibinfo{number}{2} (\bibinfo{year}{2008}),
  \bibinfo{pages}{1265--1276}.
\newblock


\bibitem[\protect\citeauthoryear{Curino, Godwal, Kroth, Kuryata, Lapinski, Liu,
  Oks, Poppe, Smiechowski, Thayer, et~al\mbox{.}}{Curino et~al\mbox{.}}{2020}]%
        {mlos}
\bibfield{author}{\bibinfo{person}{Carlo Curino}, \bibinfo{person}{Neha
  Godwal}, \bibinfo{person}{Brian Kroth}, \bibinfo{person}{Sergiy Kuryata},
  \bibinfo{person}{Greg Lapinski}, \bibinfo{person}{Siqi Liu},
  \bibinfo{person}{Slava Oks}, \bibinfo{person}{Olga Poppe},
  \bibinfo{person}{Adam Smiechowski}, \bibinfo{person}{Ed Thayer},
  {et~al\mbox{.}}} \bibinfo{year}{2020}\natexlab{}.
\newblock \showarticletitle{MLOS: An Infrastructure for Automated Software
  Performance Engineering}. In \bibinfo{booktitle}{\emph{Proceedings of the
  Fourth International Workshop on Data Management for End-to-End Machine
  Learning}}. \bibinfo{pages}{1--5}.
\newblock


\bibitem[\protect\citeauthoryear{Dash, Polyzotis, and Ailamaki}{Dash
  et~al\mbox{.}}{2011}]%
        {cophy}
\bibfield{author}{\bibinfo{person}{Debabrata Dash}, \bibinfo{person}{Neoklis
  Polyzotis}, {and} \bibinfo{person}{Anastasia Ailamaki}.}
  \bibinfo{year}{2011}\natexlab{}.
\newblock \showarticletitle{CoPhy: A Scalable, Portable, and Interactive Index
  Advisor for Large Workloads}.
\newblock \bibinfo{journal}{\emph{Proceedings of the VLDB Endowment}}
  \bibinfo{volume}{4}, \bibinfo{number}{6} (\bibinfo{year}{2011}).
\newblock


\bibitem[\protect\citeauthoryear{Ding, Das, Marcus, Wu, Chaudhuri, and
  Narasayya}{Ding et~al\mbox{.}}{2019}]%
        {ai-meets-ai}
\bibfield{author}{\bibinfo{person}{Bailu Ding}, \bibinfo{person}{Sudipto Das},
  \bibinfo{person}{Ryan Marcus}, \bibinfo{person}{Wentao Wu},
  \bibinfo{person}{Surajit Chaudhuri}, {and} \bibinfo{person}{Vivek~R
  Narasayya}.} \bibinfo{year}{2019}\natexlab{}.
\newblock \showarticletitle{Ai meets ai: Leveraging query executions to improve
  index recommendations}. In \bibinfo{booktitle}{\emph{Proceedings of the 2019
  International Conference on Management of Data}}.
  \bibinfo{pages}{1241--1258}.
\newblock


\bibitem[\protect\citeauthoryear{Foster, Gentile, Mohri, and Zimmert}{Foster
  et~al\mbox{.}}{2020}]%
        {foster2021adapting}
\bibfield{author}{\bibinfo{person}{Dylan~J Foster}, \bibinfo{person}{Claudio
  Gentile}, \bibinfo{person}{Mehryar Mohri}, {and} \bibinfo{person}{Julian
  Zimmert}.} \bibinfo{year}{2020}\natexlab{}.
\newblock \showarticletitle{Adapting to misspecification in contextual
  bandits}.
\newblock \bibinfo{journal}{\emph{Advances in Neural Information Processing
  Systems}}  \bibinfo{volume}{33} (\bibinfo{year}{2020}),
  \bibinfo{pages}{11478--11489}.
\newblock


\bibitem[\protect\citeauthoryear{Graefe}{Graefe}{1995}]%
        {cascades}
\bibfield{author}{\bibinfo{person}{Goetz Graefe}.}
  \bibinfo{year}{1995}\natexlab{}.
\newblock \showarticletitle{The cascades framework for query optimization}.
\newblock \bibinfo{journal}{\emph{IEEE Data Eng. Bull.}} \bibinfo{volume}{18},
  \bibinfo{number}{3} (\bibinfo{year}{1995}), \bibinfo{pages}{19--29}.
\newblock


\bibitem[\protect\citeauthoryear{Gupchup, Aazami, Fan, Filipi, Finley, Inglis,
  Asteborg, Caroll, Chari, Cozowicz, Gopal, Prakash, Bendapudi, Gerrits, Lau,
  Liu, Rossi, Slobodianyk, Birjukov, Cooper, Javar, Perednya, Srinivasan,
  Langford, Cutler, and Gehrke}{Gupchup et~al\mbox{.}}{2020}]%
        {gupchup2020resonance}
\bibfield{author}{\bibinfo{person}{Jayant Gupchup}, \bibinfo{person}{Ashkan
  Aazami}, \bibinfo{person}{Yaran Fan}, \bibinfo{person}{Senja Filipi},
  \bibinfo{person}{Tom Finley}, \bibinfo{person}{Scott Inglis},
  \bibinfo{person}{Marcus Asteborg}, \bibinfo{person}{Luke Caroll},
  \bibinfo{person}{Rajan Chari}, \bibinfo{person}{Markus Cozowicz},
  \bibinfo{person}{Vishak Gopal}, \bibinfo{person}{Vinod Prakash},
  \bibinfo{person}{Sasikanth Bendapudi}, \bibinfo{person}{Jack Gerrits},
  \bibinfo{person}{Eric Lau}, \bibinfo{person}{Huazhou Liu},
  \bibinfo{person}{Marco Rossi}, \bibinfo{person}{Dima Slobodianyk},
  \bibinfo{person}{Dmitri Birjukov}, \bibinfo{person}{Matty Cooper},
  \bibinfo{person}{Nilesh Javar}, \bibinfo{person}{Dmitriy Perednya},
  \bibinfo{person}{Sriram Srinivasan}, \bibinfo{person}{John Langford},
  \bibinfo{person}{Ross Cutler}, {and} \bibinfo{person}{Johannes Gehrke}.}
  \bibinfo{year}{2020}\natexlab{}.
\newblock \showarticletitle{Resonance: Replacing Software Constants with
  Context-Aware Models in Real-time Communication}. In
  \bibinfo{booktitle}{\emph{NeurIPS 2020}}.
\newblock


\bibitem[\protect\citeauthoryear{Jindal and Interlandi}{Jindal and
  Interlandi}{2021}]%
        {vldb-tutorial-2021}
\bibfield{author}{\bibinfo{person}{Alekh Jindal} {and} \bibinfo{person}{Matteo
  Interlandi}.} \bibinfo{year}{2021}\natexlab{}.
\newblock \showarticletitle{Machine Learning for Cloud Data Systems: the
  Promise, the Progress, and the Path Forward}.
\newblock \bibinfo{journal}{\emph{Proc. {VLDB} Endow.}} \bibinfo{volume}{14},
  \bibinfo{number}{12} (\bibinfo{year}{2021}), \bibinfo{pages}{3202--3205}.
\newblock
\urldef\tempurl%
\url{http://www.vldb.org/pvldb/vol14/p3202-jindal.pdf}
\showURL{%
\tempurl}


\bibitem[\protect\citeauthoryear{Jindal, Patel, Roy, Qiao, Yin, Sen, and
  Krishnan}{Jindal et~al\mbox{.}}{2019}]%
        {jindal2019peregrine}
\bibfield{author}{\bibinfo{person}{Alekh Jindal}, \bibinfo{person}{Hiren
  Patel}, \bibinfo{person}{Abhishek Roy}, \bibinfo{person}{Shi Qiao},
  \bibinfo{person}{Zhicheng Yin}, \bibinfo{person}{Rathijit Sen}, {and}
  \bibinfo{person}{Subru Krishnan}.} \bibinfo{year}{2019}\natexlab{}.
\newblock \showarticletitle{Peregrine: Workload optimization for cloud query
  engines}. In \bibinfo{booktitle}{\emph{Proceedings of the ACM Symposium on
  Cloud Computing}}. \bibinfo{pages}{416--427}.
\newblock


\bibitem[\protect\citeauthoryear{Jindal, Qiao, Patel, Yin, Di, Bag, Friedman,
  Lin, Karanasos, and Rao}{Jindal et~al\mbox{.}}{2018}]%
        {cloudviews}
\bibfield{author}{\bibinfo{person}{Alekh Jindal}, \bibinfo{person}{Shi Qiao},
  \bibinfo{person}{Hiren Patel}, \bibinfo{person}{Zhicheng Yin},
  \bibinfo{person}{Jieming Di}, \bibinfo{person}{Malay Bag},
  \bibinfo{person}{Marc Friedman}, \bibinfo{person}{Yifung Lin},
  \bibinfo{person}{Konstantinos Karanasos}, {and} \bibinfo{person}{Sriram
  Rao}.} \bibinfo{year}{2018}\natexlab{}.
\newblock \showarticletitle{Computation reuse in analytics job service at
  microsoft}. In \bibinfo{booktitle}{\emph{Proceedings of the 2018
  International Conference on Management of Data}}. \bibinfo{pages}{191--203}.
\newblock


\bibitem[\protect\citeauthoryear{Jindal, Qiao, Sen, and Patel}{Jindal
  et~al\mbox{.}}{2021}]%
        {microlearner}
\bibfield{author}{\bibinfo{person}{Alekh Jindal}, \bibinfo{person}{Shi Qiao},
  \bibinfo{person}{Rathijit Sen}, {and} \bibinfo{person}{Hiren Patel}.}
  \bibinfo{year}{2021}\natexlab{}.
\newblock \showarticletitle{Microlearner: A fine-grained Learning Optimizer for
  Big Data Workloads at Microsoft}. In \bibinfo{booktitle}{\emph{2021 IEEE 37th
  International Conference on Data Engineering (ICDE)}}. IEEE,
  \bibinfo{pages}{2423--2434}.
\newblock


\bibitem[\protect\citeauthoryear{Jyothi, Curino, Menache, Narayanamurthy,
  Tumanov, Yaniv, Mavlyutov, Goiri, Krishnan, Kulkarni, et~al\mbox{.}}{Jyothi
  et~al\mbox{.}}{2016}]%
        {jyothi2016morpheus}
\bibfield{author}{\bibinfo{person}{Sangeetha~Abdu Jyothi},
  \bibinfo{person}{Carlo Curino}, \bibinfo{person}{Ishai Menache},
  \bibinfo{person}{Shravan~Matthur Narayanamurthy}, \bibinfo{person}{Alexey
  Tumanov}, \bibinfo{person}{Jonathan Yaniv}, \bibinfo{person}{Ruslan
  Mavlyutov}, \bibinfo{person}{Inigo Goiri}, \bibinfo{person}{Subru Krishnan},
  \bibinfo{person}{Janardhan Kulkarni}, {et~al\mbox{.}}}
  \bibinfo{year}{2016}\natexlab{}.
\newblock \showarticletitle{Morpheus: Towards Automated $\{$SLOs$\}$ for
  Enterprise Clusters}. In \bibinfo{booktitle}{\emph{12th USENIX Symposium on
  Operating Systems Design and Implementation (OSDI 16)}}.
  \bibinfo{pages}{117--134}.
\newblock


\bibitem[\protect\citeauthoryear{Kraska}{Kraska}{2021}]%
        {instanceopt-datasys}
\bibfield{author}{\bibinfo{person}{Tim Kraska}.}
  \bibinfo{year}{2021}\natexlab{}.
\newblock \showarticletitle{Towards instance-optimized data systems}.
\newblock \bibinfo{journal}{\emph{Proceedings of the VLDB Endowment}}
  \bibinfo{volume}{14}, \bibinfo{number}{12} (\bibinfo{year}{2021}).
\newblock


\bibitem[\protect\citeauthoryear{Kraska, Alizadeh, Beutel, Chi, Ding, Kristo,
  Leclerc, Madden, Mao, and Nathan}{Kraska et~al\mbox{.}}{2019}]%
        {DBLP:conf/cidr/KraskaABCKLMMN19}
\bibfield{author}{\bibinfo{person}{Tim Kraska}, \bibinfo{person}{Mohammad
  Alizadeh}, \bibinfo{person}{Alex Beutel}, \bibinfo{person}{H Chi},
  \bibinfo{person}{Jialin Ding}, \bibinfo{person}{Ani Kristo},
  \bibinfo{person}{Guillaume Leclerc}, \bibinfo{person}{Samuel Madden},
  \bibinfo{person}{Hongzi Mao}, {and} \bibinfo{person}{Vikram Nathan}.}
  \bibinfo{year}{2019}\natexlab{}.
\newblock \showarticletitle{SageDB: A Learned Database System}. In
  \bibinfo{booktitle}{\emph{{CIDR}}}.
\newblock


\bibitem[\protect\citeauthoryear{Kraska, Beutel, Chi, Dean, and
  Polyzotis}{Kraska et~al\mbox{.}}{2018}]%
        {DBLP:conf/sigmod/KraskaBCDP18}
\bibfield{author}{\bibinfo{person}{Tim Kraska}, \bibinfo{person}{Alex Beutel},
  \bibinfo{person}{Ed~H Chi}, \bibinfo{person}{Jeffrey Dean}, {and}
  \bibinfo{person}{Neoklis Polyzotis}.} \bibinfo{year}{2018}\natexlab{}.
\newblock \showarticletitle{The case for learned index structures}. In
  \bibinfo{booktitle}{\emph{Proceedings of the 2018 international conference on
  management of data}}. \bibinfo{pages}{489--504}.
\newblock


\bibitem[\protect\citeauthoryear{Krishnan, Yang, Goldberg, Hellerstein, and
  Stoica}{Krishnan et~al\mbox{.}}{2018}]%
        {berkeley_drl}
\bibfield{author}{\bibinfo{person}{Sanjay Krishnan}, \bibinfo{person}{Zongheng
  Yang}, \bibinfo{person}{Ken Goldberg}, \bibinfo{person}{Joseph Hellerstein},
  {and} \bibinfo{person}{Ion Stoica}.} \bibinfo{year}{2018}\natexlab{}.
\newblock \showarticletitle{Learning to optimize join queries with deep
  reinforcement learning}.
\newblock \bibinfo{journal}{\emph{arXiv preprint arXiv:1808.03196}}
  (\bibinfo{year}{2018}).
\newblock


\bibitem[\protect\citeauthoryear{Langford and Zhang}{Langford and
  Zhang}{2007}]%
        {langford2007epoch}
\bibfield{author}{\bibinfo{person}{John Langford} {and} \bibinfo{person}{Tong
  Zhang}.} \bibinfo{year}{2007}\natexlab{}.
\newblock \showarticletitle{The epoch-greedy algorithm for multi-armed bandits
  with side information}.
\newblock \bibinfo{journal}{\emph{Advances in neural information processing
  systems}}  \bibinfo{volume}{20} (\bibinfo{year}{2007}).
\newblock


\bibitem[\protect\citeauthoryear{Leis, Gubichev, Mirchev, Boncz, Kemper, and
  Neumann}{Leis et~al\mbox{.}}{2015}]%
        {howgoodreally}
\bibfield{author}{\bibinfo{person}{Viktor Leis}, \bibinfo{person}{Andrey
  Gubichev}, \bibinfo{person}{Atanas Mirchev}, \bibinfo{person}{Peter Boncz},
  \bibinfo{person}{Alfons Kemper}, {and} \bibinfo{person}{Thomas Neumann}.}
  \bibinfo{year}{2015}\natexlab{}.
\newblock \showarticletitle{How good are query optimizers, really?}
\newblock \bibinfo{journal}{\emph{Proceedings of the VLDB Endowment}}
  \bibinfo{volume}{9}, \bibinfo{number}{3} (\bibinfo{year}{2015}),
  \bibinfo{pages}{204--215}.
\newblock


\bibitem[\protect\citeauthoryear{Marcus, Negi, Mao, Tatbul, Alizadeh, and
  Kraska}{Marcus et~al\mbox{.}}{2021}]%
        {Marcus2020BaoLT}
\bibfield{author}{\bibinfo{person}{Ryan Marcus}, \bibinfo{person}{Parimarjan
  Negi}, \bibinfo{person}{Hongzi Mao}, \bibinfo{person}{Nesime Tatbul},
  \bibinfo{person}{Mohammad Alizadeh}, {and} \bibinfo{person}{Tim Kraska}.}
  \bibinfo{year}{2021}\natexlab{}.
\newblock \showarticletitle{Bao: Making learned query optimization practical}.
  In \bibinfo{booktitle}{\emph{Proceedings of the 2021 International Conference
  on Management of Data}}. \bibinfo{pages}{1275--1288}.
\newblock


\bibitem[\protect\citeauthoryear{Marcus, Negi, Mao, Zhang, Alizadeh, Kraska,
  Papaemmanouil, and Tatbul}{Marcus et~al\mbox{.}}{2019}]%
        {marcus2019neo}
\bibfield{author}{\bibinfo{person}{Ryan Marcus}, \bibinfo{person}{Parimarjan
  Negi}, \bibinfo{person}{Hongzi Mao}, \bibinfo{person}{Chi Zhang},
  \bibinfo{person}{Mohammad Alizadeh}, \bibinfo{person}{Tim Kraska},
  \bibinfo{person}{Olga Papaemmanouil}, {and} \bibinfo{person}{Nesime Tatbul}.}
  \bibinfo{year}{2019}\natexlab{}.
\newblock \showarticletitle{Neo: a learned query optimizer}.
\newblock \bibinfo{journal}{\emph{Proceedings of the VLDB Endowment}}
  \bibinfo{volume}{12}, \bibinfo{number}{11} (\bibinfo{year}{2019}),
  \bibinfo{pages}{1705--1718}.
\newblock


\bibitem[\protect\citeauthoryear{Marcus and Papaemmanouil}{Marcus and
  Papaemmanouil}{2018}]%
        {rejoin}
\bibfield{author}{\bibinfo{person}{Ryan Marcus} {and} \bibinfo{person}{Olga
  Papaemmanouil}.} \bibinfo{year}{2018}\natexlab{}.
\newblock \showarticletitle{Deep reinforcement learning for join order
  enumeration}. In \bibinfo{booktitle}{\emph{Proceedings of the First
  International Workshop on Exploiting Artificial Intelligence Techniques for
  Data Management}}. \bibinfo{pages}{1--4}.
\newblock


\bibitem[\protect\citeauthoryear{Negi, Interlandi, Marcus, Alizadeh, Kraska,
  Friedman, and Jindal}{Negi et~al\mbox{.}}{2021}]%
        {steerqo}
\bibfield{author}{\bibinfo{person}{Parimarjan Negi}, \bibinfo{person}{Matteo
  Interlandi}, \bibinfo{person}{Ryan Marcus}, \bibinfo{person}{Mohammad
  Alizadeh}, \bibinfo{person}{Tim Kraska}, \bibinfo{person}{Marc Friedman},
  {and} \bibinfo{person}{Alekh Jindal}.} \bibinfo{year}{2021}\natexlab{}.
\newblock \showarticletitle{Steering query optimizers: A practical take on big
  data workloads}. In \bibinfo{booktitle}{\emph{Proceedings of the 2021
  International Conference on Management of Data}}.
  \bibinfo{pages}{2557--2569}.
\newblock


\bibitem[\protect\citeauthoryear{Patel, Jindal, and Szyperski}{Patel
  et~al\mbox{.}}{2019}]%
        {patel2019big}
\bibfield{author}{\bibinfo{person}{Hiren Patel}, \bibinfo{person}{Alekh
  Jindal}, {and} \bibinfo{person}{Clemens Szyperski}.}
  \bibinfo{year}{2019}\natexlab{}.
\newblock \showarticletitle{Big Data Processing at Microsoft: Hyper Scale,
  Massive Complexity, and Minimal Cost}. In
  \bibinfo{booktitle}{\emph{Proceedings of the ACM Symposium on Cloud
  Computing}}. \bibinfo{pages}{490--490}.
\newblock


\bibitem[\protect\citeauthoryear{Pavlo, Angulo, Arulraj, Lin, Lin, Ma, Menon,
  Mowry, Perron, Quah, et~al\mbox{.}}{Pavlo et~al\mbox{.}}{2017}]%
        {pavlo17}
\bibfield{author}{\bibinfo{person}{Andrew Pavlo}, \bibinfo{person}{Gustavo
  Angulo}, \bibinfo{person}{Joy Arulraj}, \bibinfo{person}{Haibin Lin},
  \bibinfo{person}{Jiexi Lin}, \bibinfo{person}{Lin Ma},
  \bibinfo{person}{Prashanth Menon}, \bibinfo{person}{Todd~C Mowry},
  \bibinfo{person}{Matthew Perron}, \bibinfo{person}{Ian Quah},
  {et~al\mbox{.}}} \bibinfo{year}{2017}\natexlab{}.
\newblock \showarticletitle{Self-Driving Database Management Systems.}. In
  \bibinfo{booktitle}{\emph{CIDR}}, Vol.~\bibinfo{volume}{4}.
  \bibinfo{pages}{1}.
\newblock


\bibitem[\protect\citeauthoryear{Pimpley, Li, Srivastava, Rohra, Zhu,
  Srinivasan, Jindal, Patel, Qiao, and Sen}{Pimpley et~al\mbox{.}}{2021}]%
        {tasq}
\bibfield{author}{\bibinfo{person}{Anish Pimpley}, \bibinfo{person}{Shuo Li},
  \bibinfo{person}{Anubha Srivastava}, \bibinfo{person}{Vishal Rohra},
  \bibinfo{person}{Yi Zhu}, \bibinfo{person}{Soundararajan Srinivasan},
  \bibinfo{person}{Alekh Jindal}, \bibinfo{person}{Hiren Patel},
  \bibinfo{person}{Shi Qiao}, {and} \bibinfo{person}{Rathijit Sen}.}
  \bibinfo{year}{2021}\natexlab{}.
\newblock \showarticletitle{Optimal resource allocation for serverless
  queries}.
\newblock \bibinfo{journal}{\emph{arXiv preprint arXiv:2107.08594}}
  (\bibinfo{year}{2021}).
\newblock


\bibitem[\protect\citeauthoryear{Power, Patel, Jindal, Leeka, Jenkins, Rys,
  Triou, Zhu, Katahanas, Talapady, et~al\mbox{.}}{Power et~al\mbox{.}}{2021}]%
        {scope2021}
\bibfield{author}{\bibinfo{person}{Conor Power}, \bibinfo{person}{Hiren Patel},
  \bibinfo{person}{Alekh Jindal}, \bibinfo{person}{Jyoti Leeka},
  \bibinfo{person}{Bob Jenkins}, \bibinfo{person}{Michael Rys},
  \bibinfo{person}{Ed Triou}, \bibinfo{person}{Dexin Zhu},
  \bibinfo{person}{Lucky Katahanas}, \bibinfo{person}{Chakrapani~Bhat
  Talapady}, {et~al\mbox{.}}} \bibinfo{year}{2021}\natexlab{}.
\newblock \showarticletitle{The cosmos big data platform at Microsoft: over a
  decade of progress and a decade to look forward}.
\newblock \bibinfo{journal}{\emph{Proceedings of the VLDB Endowment}}
  \bibinfo{volume}{14}, \bibinfo{number}{12} (\bibinfo{year}{2021}),
  \bibinfo{pages}{3148--3161}.
\newblock


\bibitem[\protect\citeauthoryear{Research}{Research}{[n.d.]}]%
        {vw}
\bibfield{author}{\bibinfo{person}{Yahoo! Research}.}
  \bibinfo{year}{[n.d.]}\natexlab{}.
\newblock \bibinfo{title}{Vowpal Wabbit}.
\newblock \bibinfo{howpublished}{\url{https://vowpalwabbit.org/research.html}}.
\newblock


\bibitem[\protect\citeauthoryear{Roy, Jindal, Gomatam, Ouyang, Gosalia, Ravi,
  Mann, and Jain}{Roy et~al\mbox{.}}{2021}]%
        {sparkcruise21}
\bibfield{author}{\bibinfo{person}{Abhishek Roy}, \bibinfo{person}{Alekh
  Jindal}, \bibinfo{person}{Priyanka Gomatam}, \bibinfo{person}{Xiating
  Ouyang}, \bibinfo{person}{Ashit Gosalia}, \bibinfo{person}{Nishkam Ravi},
  \bibinfo{person}{Swinky Mann}, {and} \bibinfo{person}{Prakhar Jain}.}
  \bibinfo{year}{2021}\natexlab{}.
\newblock \showarticletitle{SparkCruise: workload optimization in managed spark
  clusters at Microsoft}.
\newblock \bibinfo{journal}{\emph{Proceedings of the VLDB Endowment}}
  \bibinfo{volume}{14}, \bibinfo{number}{12} (\bibinfo{year}{2021}),
  \bibinfo{pages}{3122--3134}.
\newblock


\bibitem[\protect\citeauthoryear{Sen, Jindal, Patel, and Qiao}{Sen
  et~al\mbox{.}}{2020}]%
        {autotoken}
\bibfield{author}{\bibinfo{person}{Rathijit Sen}, \bibinfo{person}{Alekh
  Jindal}, \bibinfo{person}{Hiren Patel}, {and} \bibinfo{person}{Shi Qiao}.}
  \bibinfo{year}{2020}\natexlab{}.
\newblock \showarticletitle{Autotoken: Predicting peak parallelism for big data
  analytics at microsoft}.
\newblock \bibinfo{journal}{\emph{Proceedings of the VLDB Endowment}}
  \bibinfo{volume}{13}, \bibinfo{number}{12} (\bibinfo{year}{2020}),
  \bibinfo{pages}{3326--3339}.
\newblock


\bibitem[\protect\citeauthoryear{Siddiqui, Jindal, Qiao, Patel, and
  Le}{Siddiqui et~al\mbox{.}}{2020}]%
        {siddiqui2020cost}
\bibfield{author}{\bibinfo{person}{Tarique Siddiqui}, \bibinfo{person}{Alekh
  Jindal}, \bibinfo{person}{Shi Qiao}, \bibinfo{person}{Hiren Patel}, {and}
  \bibinfo{person}{Wangchao Le}.} \bibinfo{year}{2020}\natexlab{}.
\newblock \showarticletitle{Cost models for big data query processing:
  Learning, retrofitting, and our findings}. In
  \bibinfo{booktitle}{\emph{Proceedings of the 2020 ACM SIGMOD International
  Conference on Management of Data}}. \bibinfo{pages}{99--113}.
\newblock


\bibitem[\protect\citeauthoryear{Sutton and Barto}{Sutton and Barto}{2018}]%
        {sutton2018reinforcement}
\bibfield{author}{\bibinfo{person}{Richard~S Sutton} {and}
  \bibinfo{person}{Andrew~G Barto}.} \bibinfo{year}{2018}\natexlab{}.
\newblock \bibinfo{booktitle}{\emph{Reinforcement learning: An introduction}}.
\newblock \bibinfo{publisher}{MIT press}.
\newblock


\bibitem[\protect\citeauthoryear{Van~Aken, Pavlo, Gordon, and Zhang}{Van~Aken
  et~al\mbox{.}}{2017}]%
        {vanaken17}
\bibfield{author}{\bibinfo{person}{Dana Van~Aken}, \bibinfo{person}{Andrew
  Pavlo}, \bibinfo{person}{Geoffrey~J Gordon}, {and} \bibinfo{person}{Bohan
  Zhang}.} \bibinfo{year}{2017}\natexlab{}.
\newblock \showarticletitle{Automatic database management system tuning through
  large-scale machine learning}. In \bibinfo{booktitle}{\emph{Proceedings of
  the 2017 ACM international conference on management of data}}.
  \bibinfo{pages}{1009--1024}.
\newblock


\bibitem[\protect\citeauthoryear{Wang, Agarwal, and Dud{\i}k}{Wang
  et~al\mbox{.}}{2017}]%
        {wang2017optimal}
\bibfield{author}{\bibinfo{person}{Yu-Xiang Wang}, \bibinfo{person}{Alekh
  Agarwal}, {and} \bibinfo{person}{Miroslav Dud{\i}k}.}
  \bibinfo{year}{2017}\natexlab{}.
\newblock \showarticletitle{Optimal and adaptive off-policy evaluation in
  contextual bandits}. In \bibinfo{booktitle}{\emph{International Conference on
  Machine Learning}}. PMLR, \bibinfo{pages}{3589--3597}.
\newblock


\bibitem[\protect\citeauthoryear{Wu, Jindal, Amizadeh, Patel, Le, Qiao, and
  Rao}{Wu et~al\mbox{.}}{2018}]%
        {wu2018towards}
\bibfield{author}{\bibinfo{person}{Chenggang Wu}, \bibinfo{person}{Alekh
  Jindal}, \bibinfo{person}{Saeed Amizadeh}, \bibinfo{person}{Hiren Patel},
  \bibinfo{person}{Wangchao Le}, \bibinfo{person}{Shi Qiao}, {and}
  \bibinfo{person}{Sriram Rao}.} \bibinfo{year}{2018}\natexlab{}.
\newblock \showarticletitle{Towards a learning optimizer for shared clouds}.
\newblock \bibinfo{journal}{\emph{Proceedings of the VLDB Endowment}}
  \bibinfo{volume}{12}, \bibinfo{number}{3} (\bibinfo{year}{2018}),
  \bibinfo{pages}{210--222}.
\newblock


\bibitem[\protect\citeauthoryear{Wu, Zhou, Bruno, Zhang, and Fowler}{Wu
  et~al\mbox{.}}{2012}]%
        {wu2012scope}
\bibfield{author}{\bibinfo{person}{Ming-Chuan Wu}, \bibinfo{person}{Jingren
  Zhou}, \bibinfo{person}{Nicolas Bruno}, \bibinfo{person}{Yu Zhang}, {and}
  \bibinfo{person}{Jon Fowler}.} \bibinfo{year}{2012}\natexlab{}.
\newblock \showarticletitle{Scope playback: self-validation in the cloud}. In
  \bibinfo{booktitle}{\emph{Proceedings of the Fifth International Workshop on
  Testing Database Systems}}. \bibinfo{pages}{1--6}.
\newblock


\bibitem[\protect\citeauthoryear{Zhu, Interlandi, Roy, Das, Patel, Bag, Sharma,
  and Jindal}{Zhu et~al\mbox{.}}{2021}]%
        {phoebe}
\bibfield{author}{\bibinfo{person}{Yiwen Zhu}, \bibinfo{person}{Matteo
  Interlandi}, \bibinfo{person}{Abhishek Roy}, \bibinfo{person}{Krishnadhan
  Das}, \bibinfo{person}{Hiren Patel}, \bibinfo{person}{Malay Bag},
  \bibinfo{person}{Hitesh Sharma}, {and} \bibinfo{person}{Alekh Jindal}.}
  \bibinfo{year}{2021}\natexlab{}.
\newblock \showarticletitle{Phoebe: a learning-based checkpoint optimizer}.
\newblock \bibinfo{journal}{\emph{Proceedings of the VLDB Endowment}}
  \bibinfo{volume}{14}, \bibinfo{number}{11} (\bibinfo{year}{2021}),
  \bibinfo{pages}{2505--2518}.
\newblock


\end{thebibliography}

% \balance
% \pagebreak 

\end{document}